\documentclass[aps,nofootinbib,showpacs,prd,12pt]{revtex4}

\usepackage{color}
\usepackage{epsf,epsfig,amsmath,amssymb,dsfont}
\usepackage{amsthm,mathrsfs} %mathematical symbols
\usepackage{graphicx} %includes images
\usepackage{hyperref}
\usepackage[normalem]{ulem}

\DeclareMathOperator{\erf}{erf}

\newcommand{\be}{\begin{equation}}
\newcommand{\ee}{\end{equation}}
\newcommand{\baln}{\begin{align}}
\newcommand{\ealn}{\end{align}}
\newcommand{\ben}{\begin{equation*}}
\newcommand{\een}{\end{equation*}}
\newcommand{\bwt}{\begin{widetext}}
\newcommand{\ewt}{\end{widetext}}

\long\def\symbolfootnote[#1]#2{\begingroup%
\def\thefootnote{\fnsymbol{footnote}}\footnote[#1]{#2}\endgroup}

\newcommand{\nn}{\nonumber \\}

\newcommand{\fr}{\frac}
\newcommand{\del}{\partial}
\newcommand\CS{\mathcal{C}}
\newcommand\ohat{\hat{\mathcal{O}}}
\newcommand\phat{\hat{\mathcal{P}}}

\newcommand{\mbb}{\mathbb}

\newcommand{\euv}{e^{-\sigma u^2v^2}}

\newcommand{\erfone}{\fr{\erf(\sqrt{\sigma}uv)}{\sqrt{\sigma}}}
\newcommand{\erftwo}{\fr{\erf(\sqrt{\sigma}uv)}{\sigma^{3/2}}}
\begin{document}

\title{The Continuum Limit of a 4-dimensional Causal Set Scalar d'Alembertian}

\author{Alessio Belenchia${}^{1,2}$}
\author{Dionigi M. T. Benincasa${}^{1,2}$}
\author{Fay Dowker${}^{3}$}
\affiliation{
${}^1$SISSA - International School for Advanced Studies, Via Bonomea 265, 34136 Trieste, Italy,\\
${}^2$INFN, Sezione di Trieste, Trieste, Italy,\\
${}^3$Blackett Laboratory,
Imperial College, London SW7 2AZ, U.K. }

\begin{abstract}

The continuum limit 
of a 4-dimensional, discrete d'Alembertian operator for
scalar fields on causal sets is studied. The continuum limit of the mean of this operator
in the Poisson point process in 4-dimensional Minkowski spacetime is
shown to be the usual continuum 
scalar d'Alembertian $\Box$. It is shown that the mean is close to the limit when there 
exists a frame in which the 
scalar field is slowly varying on a scale set by the density of the Poisson process.
The continuum limit of the mean of the causal set d'Alembertian 
 in 4-dimensional curved spacetime is 
 shown to  equal $\Box - \frac{1}{2}R$, where $R$ is the Ricci scalar,
under certain conditions on the spacetime and the scalar field.  

\end{abstract}


\maketitle

\section{Introduction}

The existence of a physical Planck scale 
cutoff is indicated from many different directions in physics, most notably and convincingly by the value of 
the black hole entropy \cite{Sorkin:1997gi}. Perhaps the simplest response to this is to postulate
that spacetime is fundamentally discrete at the Planck scale. 
Causal sets are discrete spacetimes proposed as the histories in a sum-over-histories approach to 
quantum gravity which embody the breakdown of continuum spacetime at the Planck scale 
whilst preserving Lorentz symmetry ~\cite{Bombelli:1987aa}.  
Even if one believes that some other 
substance -- strings or loops or something else -- will turn out to 
be more relevant to the physics of quantum gravity at the Planck scale, causal sets can be useful as models of spacetime with no structure on scales smaller than 
the Planck scale which respect
a physical symmetry that has been the basis for enormous progress in fundamental physics.

That discreteness can be compatible with Lorentz invariance
is welcome news for workers  guided by the unity of physics. However, there is 
a price: the discreteness and Lorentz invariance of causal 
sets together result in a radical nonlocality \cite{Bombelli:1987aa,Moore:1988zz,Bombelli:1988qh}.
This nonlocality, were it incorrigible, could prevent causal sets from 
being useful phenomenologically and would 
threaten to derail the causal set programme for quantum gravity.
Thus,  evidence that the nonlocality of Lorentz invariant discrete structure can
be tamed \cite{Daughton:1993,Salgado:2008} is important to the causal set programme. 
More recently further evidence was provided by the discovery of 
a quasi-local, discrete scalar d'Alembertian operator, 
$B^{(2)}$ for fields on causal sets well-approximated by $2$ 
dimensional Minkowski spacetime \cite{Sorkin:2007qi,Henson:2006kf}. The mean of this operator
tends to the exact continuum 2 dimensional flat scalar d'Alembertian in the 
continuum limit \cite{Sorkinnotes}. 
In \cite{Benincasa:2010ac} this work was extended to 
4 dimensions with the introduction of an analogous operator $B^{(4)}$.
There it was claimed that for both $d=2$ and $d=4$, when
$B^{(d)}$ is 
applied to scalar fields on causal sets which are approximated 
by $d$-dimensional Lorentzian spacetimes, its mean tends
in the continuum limit to the 
curved space operator, $\Box - {\frac{1}{2}} R$, where $\Box$ is the curved spacetime
scalar d'Alembertian and $R$ is the Ricci  scalar curvature. 
In this paper we will prove this result in four dimensions
under certain conditions. 
Note that the equation of motion $B^{(d)} \phi = 0$ for the field on the causal set results in a non-minimal coupling to 
gravity in the continuum limit in all dimensions 
dimensions~\cite{Dowker:2013vl,Glaser:2013sf,0264-9381-33-13-135011}.
% and may have interesting connections with Einstein's 
%Equivalence Principle (see~\cite{0264-9381-33-13-135011}).

Although these operators (and their generalisations to any dimension \cite{Dowker:2013vl, Glaser:2013sf, Aslanbeigi:2014tg}) 
do indeed tame the radical nonlocality referred to above,
they do not eliminate it altogether. This remnant of nonlocality
is manifest in the dynamics of (scalar) fields on 
spacetime, which becomes nonlocal.  Nonlocal dynamics of exactly this form
was recently used in the construction of scalar nonlocal quantum field theories 
in \cite{Belenchia:2014fda} and \cite{Saravani:2015aa}, potentially 
leading to novel and interesting phenomenology~\cite{Saravani:2016aa,Belenchia:2016aa}.

Recall that a causal set (or causet) is a locally finite partial order, $(\CS,\preceq)$. 
Local finiteness is the condition that 
the cardinality of any {\emph{order interval}} is finite, 
where the (inclusive) order interval between a pair of elements
$y\preceq x$ is defined to be
$I(x,y) := \{z\in \CS \,|\, y\preceq z \preceq x\}$.
We write $x \prec y$ when $x \preceq y$ and $x \ne y$.
We call a relation $x\prec y$ a \emph{link} 
if the order interval $I(x,y)$ contains only $x$ and $y$. We denote by $|\cdot|$ the cardinality 
of a set and $n(x,y):=|I(x,y)|-2$. 
 
Given a point $x\in\CS$ we define the set of all its past nearest neighbours 
to be
\be L_1(x):=\{y\in \CS\;|\;y\prec x,\; n(x,y) = 0\}.
\label{2.2}\ee
We refer to this set of elements as the first \textit{past layer}. We can 
generalise this by defining the sets of next nearest neighbours, $L_2$,
next next nearest neighbours, $L_3$, and so on. In general the $i$-th
past layer is defined as
\be L_i(x):=\{y\in \CS \;|\; y\prec x \; \text{and} \; n(x,y)=i-1\}
\label{2.3}.\ee

Consider the discrete retarded operator
$B$, on a causet $\CS$, defined as follows
\cite{Benincasa:2010ac}. If $\phi:\CS\rightarrow\mbb{R}$ is a scalar field, then
\begin{align}
B\phi(x):=\frac{4}{\sqrt{6}l^2}\Bigg[-\phi(x)
+(\sum_{y \in L_1(x)}-
9\sum_{y\in L_2(x)}+16\sum_{y\in L_3(x)}-8\sum_{y\in L_4(x)})\phi(y)\Bigg]\,,
\label{Bop}
\end{align}
where 
$l$ is a length (the analogue of the lattice spacing). 
The form of the discrete operator $B$ 
as such a sum-over-layers is dictated by requiring the operator to be a difference operator analogous to the d'Alembertian on a lattice but in addition requiring it to be 
retarded and Lorentz invariant (see~\cite{Sorkin:2007qi} for an explanation in the 
2 dimensional case). Four is the minimum number of layers in four dimensions and the specific values of the coefficients are required to give the correct local limit in Minkowski space. Introducing more layers is possible but is analogous to, say, approximating a second derivative in one dimension by a sum over values of the function at more than three neighbouring points: nonuniqueness of the coefficients results~\cite{Aslanbeigi:2014tg}.

 $B$ is defined on scalar fields on any causal set but 
 is physically relevant for causal sets that are well-approximated 
 by a four dimensional Lorentzian manifold, $(\mathcal{M},g)$.
 A causet, $(\CS, \preceq)$ is well approximated by $(\mathcal{M},g)$
 if there exists a {\textit{faithful embedding}}
 of $\CS$ into $\mathcal{M}$ in which the 
 causal order of the embedded elements respects the order of $\CS$
 and in which the number of causet elements embedded in any 
 sufficiently nice, large region of $\mathcal{M}$ approximates the spacetime volume of that 
 region in fundamental units.  These manifold-like, faithfully embeddable  causets 
  are typical in the random process of 
 \emph{sprinkling} into $(\mathcal{M},g)$: 
 a Poisson process of selecting points in $\mathcal{M}$
with density $\rho$ so that the  expected number of points 
sprinkled in a region of spacetime volume $V$ is $\rho V$. 
To do justice to our expectations for quantum gravity, the 
density $\rho = l^{-4}$, where $l$ is the fundamental length scale of the 
order of the Planck length. 
The probability for sprinkling $m$ elements into a region of volume V is
\be P(m) = \fr{(\rho V)^me^{-\rho V}}{m!}\label{2.1},\ee
This process generates 
a causet, $\CS$ whose elements are the sprinkled points and whose 
order, $\preceq$  is that induced by the manifold's causal 
order restricted to the sprinkled points. 

Let $\phi$ be a real 
test field of compact support on $\mathcal{M}$ and
$x\in \mathcal{M}$. If we sprinkle $\mathcal{M}$
at density $\rho$,
include $x$ in the resulting causet, $\CS$, 
then  $L_1(x)\subset \CS$ will be a set whose 
elements lie in the causal past of $x$, $J^-(x)$, and hug 
the boundary of  $J^-(x)$. Their locus is roughly 
the hyperboloid which lies one Planck unit of proper time to the past of $x$. 
The elements of $L_2(x)$ will also be distributed down the 
inside of the boundary of $J^-(x)$, just inside layer 1, and 
so on. The operator $B$ can be 
applied, at point $x$,  to the field $\phi$ restricted to the sprinkled causet:
 $B\phi(x)$ looks highly nonlocal, involving the value of $\phi$ at
enormous numbers of points outside any fixed neighbourhood 
of $x$. 

The sprinkling process at density $\rho$ produces, for each point 
$x$ of $\mathcal{M}$,  a random variable whose value is
$B\phi(x)$ on the realisation $\CS$ of the process. 
The expectation value of  this random variable is given by the spacetime integral 
\begin{align}
\bar{B}\phi(x):=
\mbb{E}(B\phi(x)) = 
\fr{4\sqrt{\rho}}{\sqrt{6}}\big[-\phi(x)
+\rho\int_{y\in J^-(x)}
\!\!\!\!\!\!\!d^4y\;\sqrt{-g}\; \phi(y)\,
e^{-\xi}(1-9\xi+8\xi^2-\frac{4}{3}\xi^3)\big],
\label{Bbar}
\end{align}
where $\xi:=\rho V(y)$ and $V(y)$ is the volume of the 
causal interval between $x$ and $y$. 

We can see that the integrand is suppressed wherever
$\xi$ is large, {\it{i.e.}} wherever the spacetime volume 
of the causal interval between $x$ and $y$ is larger than a few
Planck volumes. However, $\xi$ is small in the
part of the region of integration  close
to the boundary of $J^-(x)$ and, by itself, the exponential factor in the 
integrand cannot provide the suppression required to give 
a value that is approximately a local quantity at $x$. 
In the following we will show that, for large enough $\rho$,  $\bar{B}\phi(x)$ 
is effectively local and is dominated by contributions from a neighbourhood of
$x$: the contributions from far down the boundary of $J^-(x)$ 
cancel out. Indeed, we will show
\be  \label{lim}
\lim_{\rho\rightarrow \infty}\bar{B}\phi(x) = \Box\phi(x)-\frac{1}{2}R(x)\phi(x)
\ee
under certain assumptions about the support of $\phi$ in $(\mathcal{M}^{(4)},g)$.
We use the conventions of Hawking and Ellis \cite{hawking1973large}.

\section{Minkowski Spacetime}\label{flatsection}

First we consider the simpler
case of a sprinkling in Minkowski spacetime for which we will 
show that $\lim_{\rho\rightarrow \infty}\bar{B}\phi(x) = \Box\phi(x)$. Although this is
 a special case of the curved space result, it is useful to see this simpler proof
 first as it will provide the basic framework on which the curved spacetime calculation is built.
 We will also be able to estimate the corrections to the limiting value, something that will
turn out to be  harder  in the curved case. 

Choose $x$ as the origin of cartesian coordinates $\{y^\mu\}$ and in that frame
define the usual spatial polar coordinates: $r = \sqrt{\sum_{i= 1}^3 (y^i)^2}$, $\theta$ and $\varphi$. 
Null radial coordinates (past pointing) are defined by $u = \frac{1}{\sqrt{2}}(-t - r)$
and $v =  \frac{1}{\sqrt{2}}(-t + r)$ where $t = y^0$.  The volume, $V(y)$, of the causal interval between
point $y$ with cartesian coordinates $\{y^\mu\}$ and the origin is $V(y) = \frac{\pi}{6} u^2 v^2$. 

\begin{figure}[h]
\begin{center}
\includegraphics[scale=0.4]{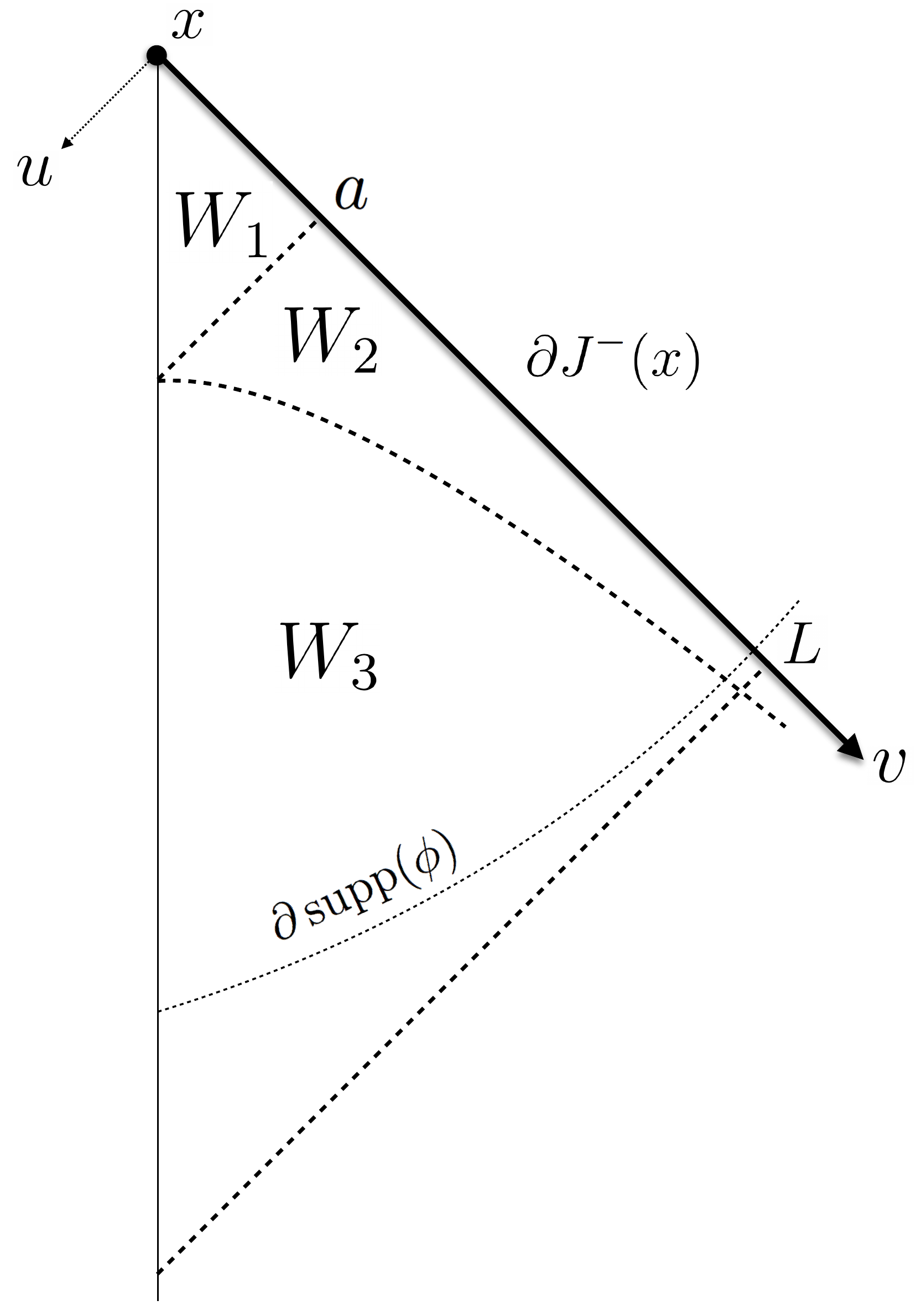}
\caption{Partition of $\mathcal{W}$ into regions $W_1$, $W_2$ and $W_3$ in $(t$-$r)$--plane}
\label{minkregions}
\end{center}
\end{figure}

Let us take the region of
integration ${\mathcal{W}}$ to be the portion of the causal past of the origin for which 
$v \le L$, where $L$ is large enough that the support of 
$\phi$ is contained in  ${\mathcal{W}}$. 
${\mathcal{W}}$ can be split into 3 parts:
\begin{align}
W_1 &:= \{y \in {\mathcal{W}}  \,|\, 0\le u\le v\le a \} \\
W_2 &:= \{y \in {\mathcal{W}} \,|\, a\le v\le L,\; 0\le u \le \frac{a^2}{v}\} \\
W_3 &:= {\mathcal{W}} \setminus (W_1 \cup W_2)\,,
\end{align}
where
 $a>0$ is chosen small enough that the expansions of $\phi$ used in the following 
 calculation are valid.  $W_1$ is a neighbourhood of the origin,
  $W_2$ is a neighbourhood
 of the past light cone and bounded away from the origin and $W_3$ is 
 a subset of the interior of the causal past that is bounded away from the 
 light cone, see Figure \ref{minkregions}. 
 
Let 
\begin{align} I_i =
\int_{W_i}
                 %&
                  d^4y\;
                              \phi(y) e^{-\rho V(y)}%\nn
%&\times 
(1-9\rho V(y)+8(\rho V(y))^2-\frac{4}{3}(\rho V(y))^3)\,,
\end{align}
for $ i = 1,2,3$ so that 
\begin{equation}
\bar{B}\phi(x):=
\fr{4\sqrt{\rho}}{\sqrt{6}}\big[-\phi(x)
+\rho(I_1 + I_2 + I_3)\big]\,.
\label{BbarJ}
\end{equation}

\subsection{The ``deep chronological past," $W_3$}

We first consider first $I_3$.
$V(y)$ is bounded away from zero in $W_3$, 
indeed $V(y) \ge V_{\textrm{min}} = \frac{\pi}{6} a^4 $ so
\begin{widetext}
\begin{equation}\left\vert I_3 \right\vert 
 \le  e^{-\rho V_{\textrm{min}} }\int_{W_3} d^4 y
 \left\vert \phi(y)
(1-9\rho V(y)+8(\rho V(y))^2-\frac{4}{3}(\rho V(y))^3) \right\vert 
\end{equation}
\end{widetext}
which tends to zero faster than any power of $\rho^{-1}$ as $\rho \rightarrow \infty$. In what follows, 
we will often write ``up to exponentially small terms'', which means we are neglecting terms
like $I_3$.

\subsection{Down the light cone, $W_2$}
Consider now $I_2$.
Note first that 
\be
e^{-\xi} 
(1-9\xi+8\xi^2-\frac{4}{3}\xi^3) 
= \ohat  e^{-\xi}, 
\ee
where 
\begin{align} 
\ohat :&= \frac{4}{3}(H + \frac{1}{2})(H + 1)(H+ \frac{3}{2})\\ 
&= 1 + 9H_1 + 8H_2 + \frac{4}{3} H_3
\end{align}
and
\be\label{hom}
H_n:= \rho^n\frac{\partial^n}{\partial \rho^n}\ \ \ {\textrm{and}}\ \ \ H:=H_1\,.
\ee
The integral we are evaluating can be rewritten as 
\begin{equation} \label{i2flat}
 I_2 =\int d\Omega_2\, \ohat \int_a^L dv \int_0^{\frac{a^2}{v}} 
du \, \frac{1}{2}(v-u)^2 \phi(y) e^{-\sigma u^2v^2}\,,
\end{equation}
where $\int d\Omega_2 = \int_0^\pi d\theta \sin{\theta} \int_0^2\pi d\varphi$. 
We have absorbed a factor of $\pi/6$ into $\sigma = \pi\rho/6$, 
for convenience. Note that the form of $\ohat$ is 
unchanged by that manoeuvre: 
\begin{equation} \label{ohatagain}
H = \rho \frac{\partial}{\partial \rho} = \sigma \frac{\partial}{\partial \sigma}\,.
\end{equation}
We will see that $I_2 = O(\rho^{-2})$ (equivalently $O(\sigma^{-2})$) and 
so, when multiplied by $\rho^{3/2}$, makes no contribution to the limit of $\bar{B}\phi(x)$. 
This can be understood by
noticing that $\ohat$ annihilates 
$\rho^{-1/2}$, $\rho^{-1}$ and $\rho^{-3/2}$. 
If the function of $\rho$ on which $\ohat$ acts is well-behaved enough as
$\rho \rightarrow \infty$ to be equal to a power series expansion in $\rho^{-1/2}$,
then application of $\ohat$ will eliminate all the terms that would not -- after multiplication by
$\rho^{3/2}$ -- tend to zero. 
Another way to understand the result is to notice that the integral over $W_2$ involves an integral
over the null coordinate $u$, transverse to the light cone. If the range of the $u$ integration 
is small enough -- if $W_2$ is close enough to the light cone -- then $\phi$ will 
be approximately constant in $u$ at fixed values of the other coordinates. The integration over
$u$ for \textit{constant} $\phi$ is  
\begin{equation} {} \int_0^{\frac{a^2}{v}} du\, \left(1 - 9\sigma v^2 u^2 + 8(\sigma v^2 u^2)^2 - \frac{4}{3}(\sigma v^2 u^2)^3\right)
(v-u)^2 e^{-\sigma v^2 u^2} \,.\label{uintegration}  
\end{equation}
The value of this integral is exponentially suppressed by 
a factor of $\exp(-\sigma a^4)$. This suggests that the 
leading, finite $\rho$  corrections to the limit are
set by the $u$ derivatives of $\phi$ at $u=0$, and this turns out to be the case.  

We assume that $a$ is chosen small enough and that $\phi$ is differentiable 
enough that throughout $W_2$, 
$\phi(y)$ can be expanded in the transverse coordinate $u$:
\begin{equation}
\phi(y) = \phi|_{u=0} + u\phi_{,u}|_{u=0} +\frac{1}{2!} u^2\phi_{,uu}|_{u=0} +\frac{1}{3!} u^3\Phi(y)\,,
\end{equation}
where $\Phi(y)$ is a continuous function. Let 
$I_2 = I_{2,0}+I_{2,1}+I_{2,2}+I_{2,3}$ where 
\begin{align}\label{j2i}
I_{2,i} = &\,\int d\Omega_2 \int_a^L dv\, F_i\int_0^{\frac{a^2}{v}}du \frac{(v-u)^2}{2}\,u^i(1-9\sigma u^2v^2+8\sigma^2u^4v^4
-\fr{4}{3}\sigma^3u^6v^6)e^{-\sigma u^2v^2},
\end{align}
and 
$F_i =\phi|_{u=0}$, $\phi_{,u}|_{u=0}$ and $\fr{1}{2}\phi_{,uu}|_{u=0}$ for $i=0,1$ and $2$ respectively, and
\begin{align}\label{j23}
I_{2,3} = &\,\int d\Omega_2 \int_a^L dv\int_0^{\frac{a^2}{v}} du\, \frac{\Phi(y)}{3!}\fr{(v-u)^2}{2}\,u^3
(1-9\sigma u^2v^2+8\sigma^2u^4v^4-\fr{4}{3}\sigma^3u^6v^6)e^{-\sigma u^2v^2}.
\end{align}

For $I_{2,i}$, $i = 0,1,2$, the $u$ and $v$ integrations can be done explicitly and the 
remaining integral over the angular coordinates can be bounded by bounding
$F_i$ by its uniform norm over the light cone $u=0$. 
We find that, up to exponentially small contributions, $I_{2,0}$ 
vanishes and 
 \begin{align}
|I_{2,1}|&\le \fr{\pi \|\phi_{,u}\|_{LC}}{3{\sigma}^2}\left(\fr{1}{a^3}-\fr{1}{L^3}\right)\\
|I_{2,2}|&\le \fr{\pi\|\phi_{,uu}\|_{LC}}{{2\sigma}^2}\left(\fr{1}{a^2}-\fr{1}{L^2}\right)+O\left(\fr{1}{\sigma^{5/2}}\right),
\end{align}
where $\| \cdot \|_{LC}$ denotes the uniform norm over the light cone $u=0$ in $W_2$. 

Finally we must bound $I_{2,3}$: 
 \begin{align}
|I_{2,3}| &\le \frac{1}{3!} \|\Phi\|_2  \int d\Omega_2 \int_a^L dv\int_0^{\frac{a^2}{v}} du\, \fr{(v-u)^2}{2}\,u^3
(1+9\sigma u^2v^2+8\sigma^2u^4v^4+\fr{4}{3}\sigma^3u^6v^6)e^{-\sigma u^2v^2}\nonumber\\
{}&=  \frac{4\pi}{12} \|\Phi\|_2 \, (1 - 9 H_1 + 8 H_2 - \frac{4}{3} H_3) \int_a^L dv\int_0^{\frac{a^2}{v}} du\, (v-u)^2 \, u^3 e^{-\sigma u^2v^2}\,,
\end{align}
where $\|\Phi\|_2$ is the uniform norm of $\Phi$ in $W_2$ and, 
is less than or equal to the uniform norm of the 
third $u$-derivative of $\phi$ in $W_2$. 
We find
\be
|I_{2,3}| \le \fr{ 33 \pi \|\phi_{,uuu}\|_2}{{ 2 \sigma}^2}\left(\fr{1}{a}-\fr{1}{L}\right)+O\left(\fr{1}{\sigma^{5/2}}\right)\,,
\ee
where $\| \cdot \|_{2}$ is the uniform norm in $W_2$. 
The key here is that $\phi$ has been expanded in $u$ far enough 
that the power of $u$ in the factor $u^3$ in $(\ref{j23})$ 
is high enough for the $u$ integration to bring down enough powers of $\sigma^{-1}$.
We will see 
the same thing happening in the integral over region $W_1$ and again in the 
curved space case.

Multiplying $I_2$ by $\rho^{3/2}$, we see that the contribution to $\bar{B}\phi(x)$ 
from the region $W_2$ tends to zero in the limit and
the leading corrections go like $\rho^{-1/2}$ and are proportional to the $u$-derivatives of 
$\phi$ on and close to the light cone. 

\subsection{The near region, $W_1$}\label{flatnear}

Now consider  
 \begin{align} 
 I_1  =&\, \ohat \int_{W_1} d^4y\, \phi(y) e^{-\rho V(y)}\\
 = &\, \ohat \int_0^a dv \int_0^v 
du \int d\Omega_2\, \frac{1}{2}(v-u)^2 \phi(y) e^{-\rho V(y)}\,. \label{i1flat}
\end{align}
We assume we can expand the field in $W_1$,
\be \label{phiexp}
\phi(y) = \phi(0) + y^{\mu}\phi_{,\mu}(0) + \fr{1}{2}y^{\mu}y^{\nu} \phi_{,\mu\nu}(0)
+ y^{\mu}y^{\nu}y^{\alpha} \psi_{\mu\nu\alpha}(y),
\ee
where $\psi_{\mu\nu\alpha}(y)$ are $C^3$-functions of the $y^{\mu}$ (they are not components 
of a tensor, the indices just label the functions).
The first two terms of the above expansion of $\phi$ contribute to $I_1$
\begin{align}
&\ohat \int_{0}^a dv \int_{0}^v du \int d\Omega_2
\fr{(v-u)^2}{2} \phi(0) e^{-\rho \fr{\pi}{6}u^2v^2} \\
&=  \frac{1}{\rho}(1-e^{-\frac{\pi}{6} \rho a^4})\phi(0),\label{canceldelta}
\end{align}
and
\begin{align}
&\ohat \int_{0}^a dv \int_{0}^v du \int d\Omega_2
\fr{(v-u)^2}{2} y^{\mu} \phi_{,\mu}(0) e^{-\rho \fr{\pi}{6}u^2v^2} \\
&= \left( -\frac{6 \sqrt{2}}{a^3 \pi\rho ^2}
+\left(\frac{\sqrt{2} a }{\rho} +\frac{6 \sqrt{2}}{a^3 \pi\rho ^2}\right)e^{-\frac{\pi}{6} \rho a^4 }\right)\phi_{,t}(0)\,.\nn
\end{align}
The first term (\ref{canceldelta}) cancels with the term $\phi(x)$ in the expression for 
$\bar{B}(x)$ (\ref{Bbar}),  while the second contributes nothing in the
limit $\rho\rightarrow\infty$. The leading correction at finite $\rho$  behaves as
\begin{equation}
\fr{l^2}{a^3} \phi_{,t}(0)\,.
\end{equation}

The third term of the expansion of $\phi$ (\ref{phiexp}) is of most 
interest to us: it contributes to $\bar{B}(x)$
\begin{align}
\frac{4}{\sqrt{6}}\rho^{\fr{3}{2}}&\ohat \int_{0}^a dv \int_{0}^v du \int d\Omega_2
\fr{1}{2}(v-u)^2 y^{\mu}y^{\nu} \phi_{,\mu\nu}(0) e^{-\rho \fr{\pi}{6}u^2v^2} \\
&=\Box\phi(0) - \frac{4\sqrt{6}}{a^2\pi\sqrt{\rho}}\phi_{,ii}(0)
+\frac{9}{a^4 \pi \rho}\left( \phi_{,ii}(0)
+ 3\phi_{,tt}(0)\right)
\label{box}
\end{align}
up to exponentially small terms (there is a sum on $i$ implied in $\phi_{,ii}$).
The leading correction at finite $\rho$ is, up to a factor of order one, 
\begin{equation}
\fr{l^2}{a^2} \phi_{,ii}(0)\,.
\end{equation}

Finally we need to show that the integral
\be
\ohat \int_{0}^a dv \int_{0}^v du \int d\Omega_2
\fr{(v-u)^2}{2} y^{\mu}y^{\nu}y^{\alpha} \psi_{\mu\nu\alpha}(y) e^{-\rho \fr{\pi}{6}u^2v^2} 
\label{nearint}
\ee
does not contribute in the limit, where $y^\mu =(t,r\cos\theta, r\sin\theta\cos\varphi, r\sin\theta\sin\varphi)$.
% \textcolor[rgb]{1,0,0}{\textbf{Here $y^{0}=t$ whereas the remaining Cartesian coordinates can be expressed in terms of spherical ones as}} 
%Each $y^\mu$ is equal to 
%$t$,} $r\cos\theta$, $r\sin\theta\cos\varphi$ or $r\sin\theta\sin\varphi$. 
At the end of the calculation,  the uniform norm of the integrand over the region of integration will be used
to bound the integral, and 
the angular dependent factors of $\cos\theta$ \textit{etc.} will make no difference to the result
and we can drop them now, for convenience. We therefore need to show that each integral of the form 
\be
\ohat \int d\Omega_2 
\int_0^adv\int_0^v du\,(v-u)^2u^mv^n\,{\psi}\, e^{-\sigma u^2v^2},\quad  m+n=3\,,
\label{nearcorrK}
\ee 
tends to zero faster than $\rho^{-3/2}$, where $\psi$ stands for one of the $ \psi_{\mu\nu\alpha}(y)$ 
and is a function of $u, v, \theta$ and $\phi$, and, again, for convenience we have defined 
$\sigma = \pi \rho/6$.

Leaving the integration over
the angles for later, consider
\be
K_{m,n} : =\ohat\int_0^adv\int_0^v du\,(v-u)^2u^mv^n\,{\psi}\, e^{-\sigma u^2v^2},\quad  m+n=3.
\label{nearcorr}
\ee 
Note first that 
\be
\euv =\fr{\sqrt{\pi}}{2v} \fr{\del}{\del u}\erfone.
\label{eerf}
\ee
Using this identity and integrating (\ref{nearcorr}) by parts in $u$ we find
\be
K_{m,n} = -\ohat \int_0^adv\int_0^v du\,\fr{\del}{\del u}\left((v-u)^2u^mv^n\psi\right)\fr{\sqrt{\pi}}{2v} \erfone,
\ee
since the boundary terms vanish.
The following identity 
\begin{equation}\label{pat2}
\ohat \left(\fr{\erf(\sqrt{\sigma}uv)}{\sqrt{\sigma}}\right)=\fr{2}{\sqrt{\pi}}uv
\phat e^{-\sigma u^2v^2},
\end{equation}
where $\phat =\fr{2}{3}(H+1)(H+\fr{3}{2})$ allows one to rewrite the integral as
\be
K_{m,n} = -\phat \int_0^adv\int_0^v du\,\fr{\del}{\del u}\left((v-u)^2u^mv^n\psi\right)u\, \euv.
\ee
Integrating by parts again in $u$ we find
\be
-\phat \int_0^adv \int_0^v du\,\fr{\del^2}{\del u^2}\left((v-u)^2u^mv^n\psi\right)\fr{\euv}{2\sigma v^2}.
\ee

Using (\ref{eerf}) and integrating by parts in $u$ again we find
\be
K_{m,n}= \phat \int_0^adv\left\{-\frac{\sqrt{\pi}}{2} \frac{\erf(\sqrt{\sigma} v^2)}{\sigma^{3/2}}v^{m+n}\tilde\psi+\int_0^v du\,\fr{\sqrt{\pi}}{4v^3}\fr{\del^3}{\del u^3}
\left((v-u)^2u^mv^n\psi\right)\erftwo  
\right\}.
\ee
where $\tilde\psi := \tilde\psi(v, \theta, \phi) = \psi|_{u = v}$. 
Using the following identity
\be
\phat\left(\fr{\erf(\sqrt{\sigma}z)}{\sigma^{3/2}}\right)=-\fr{2}{3\sqrt{\pi}}
z^3e^{-\sigma z^2},
\label{pat}
\ee
gives
\be
K_{m,n}= -\fr{1}{6}\int_0^adv\int_0^vdu\,u^3\fr{\del^3}{\del u^3}
\left((v-u)^2u^mv^n\psi\right)\euv + \fr{1}{3}\int_0^a dv\,\tilde\psi\, v^{6+m+n}e^{-\sigma v^4}\,.\label{Kmn}
\ee

Including now the integration over angles, the contribution of the second term of (\ref{Kmn})  to $I_1$, is bounded by 
\be \int d \Omega_2
\Big|\fr{1}{3}\int_0^a dv\,\tilde\psi \,v^{6+m+n}e^{-\sigma v^4}\Big|\le  4 \pi \fr{\|\tilde \psi\|_1}{{\sigma}^{5/2}}\fr{\Gamma(5/2)}{12}
\label{bndrycorr}
\ee
up to exponentially small terms,
where $\|.\|_1$ is the uniform norm over region $W_1$, we have used that $m+n=3$, and we recall that
 $\sigma = \pi\rho/6$. 
When multiplied by $\rho^{3/2}$ this term therefore gives a correction of $O(\rho^{-1})$.

Consider now the first integral in \eqref{Kmn}, and denote  a term
in the expansion of $(v-u)^2u^mv^n$ by $u^iv^j$, $i+j=5$,.
Then 
\be
u^3\fr{\del^3}{\del u^3}\left(u^iv^j\psi\right)=i(i-1)(i-2)u^{i}v^j\psi + 3i(i-1)u^{i+1}v^j\psi'+3iu^{i+2}v^j\psi''+u^{i+3}v^j\psi''',
\ee
where $'$ denotes differentiation with respect to $u$.
We can write any such term as $u^{i+k}v^j\psi^{(k)}$ where $\psi^{(n)} = \frac{\partial^n}{\partial u^n}\psi$,
$i+k\ge 3$ and $k=0,1,2,3$.

Then,
the contribution of each of these terms to
$I_1$ is bounded -- up to exponentially small terms --  by 
$4 \pi$ (from the integration over the angles) times
\be
\fr{\|\psi^{(k)}\|_1}{2(i+k-j)}\left(\fr{1}{{\sigma}^{\fr{i+j+k+2}{4}}}\Gamma\left(\fr{i+j+k+2}{4}\right)-\fr{a^{j-i-k}}{{\sigma}^{\fr{i+k+1}{2}}}
\Gamma\left(\fr{i+k+1}{2}\right)\right)
\label{bulkcorr}
\ee
for $i+k\ne j$, and
\be
\fr{\|\psi^{(k)}\|_1}{8 \sigma^{\fr{i+k+1}{2}}}\Gamma\left(\fr{i+k+1}{2}\right) 
\left( \log(\sigma a^4) - \Psi\left(\fr{i+k+1}{2}\right)  \right) \ee
for $i+k=j$ where $\Psi(z)=d\ln\Gamma(z)/dz$ is the Euler $\Psi$-function. 
%(see 8.36 and 9.3 in  \cite{gradshteyn2007}).

All these terms tend to zero in the limit and the leading order, finite $\rho$ correction
occurs when $i+j=5$, $k=0$ and (after being multiplied by $\rho^{3/2}$) 
is  $O( \rho^{-1/4}) $. 

All contributions to 
$\bar{B}\phi(x)$  have now been accounted for and we see that its limit 
is $\Box\phi(x)$ as $\rho \rightarrow \infty$.

\subsection{Finite $\rho$ corrections}

The correct continuum value for the limit of the mean is a good sign. 
However, for causal sets the important question is how $\bar{B}\phi(x)$ behaves
 when the discreteness length $l$ is 
of order the Planck length so that  $\rho = l^{-4}$ is large but finite.

The above calculations show that  
$\bar{B}\phi$ is a good approximation to $\Box\phi$ at finite $\rho$ whenever
there exists a coordinate frame and a length scale $a$ such that 
$\rho a^4 \gg 1$, \textit{i.e.} $l \ll a$ --- so that the ``exponentially small terms" 
referred to in the calculations are indeed small --- 
and such that the following conditions on the derivatives of $\phi$
in that frame hold.

The leading order corrections from  $W_1$, the neighbourhood around the origin of size $a$, give conditions
\begin{align}
&\fr{l^2}{a^3} \phi_{,t}(0)\,,\;
\fr{l^2}{a^2} \phi_{,ii}(0)\,\ll \, \Box\phi(0),\quad \quad \textrm{and}\\
& l \|\psi\|_1    \,,\;
l^2a\|\psi^{(2)}\|_1  \,,\;
l^2 a^2 \|\psi^{(3)}\|_1  \,,\;
l^2\log\left(\frac{a}{l}\right)\|\psi^{(1)}\|_1\,,\; l^4\log\left(\frac{a}{l}\right)\|\psi^{(3)}\|_1\;
\ll \, \Box\phi(0)\,,
\label{near-corr}
\end{align}
where $\psi^{(k)}$ denotes the $k$-th derivative of $\psi$ with respect to $u$. 
Recall that $\psi$ stands for a third derivative of the field $\phi$ with respect to RNC and so 
a term like $\psi^{(3)}$ is a sixth derivative of the field in RNC in the neighbourhood of 
the origin. We see that, if $l \ll a$ and $a < \lambda$, where $\lambda$ is the characteristic 
scale on which $\phi$ varies, then these conditions hold.

The leading order corrections from $W_2$, close to the light cone, give conditions
\begin{align}
\fr{l^2}{a^3} \|\phi_{,u}\|_{LC}\,,\;
\fr{l^2}{a^2} \|\phi_{,uu}\|_{LC}\,,\;
\fr{l^2}{a} \|\phi_{,uuu}\|_{2}\,\ll\, \Box\phi(0)\,.
\label{dlc-corr}
\end{align}
Note that these conditions apply to derivatives of the field $\phi$ with respect to $u$ 
on the light cone and in its neighbourhood $W_2$. If $l \ll a$ and $a < \lambda_u$ 
then, these conditions will be satisfied. 

The conclusion is that if a frame and a scale $a \gg l $ exist for which $\phi$ is slowly varying
on the scale of $a$ in a neighbourhood of $x$  and is slowly varying on that scale 
transverse to the past light cone of $x$, in a neighbourhood of the light cone, 
then $\bar{B}\phi$ is a good approximation to $\Box\phi$. 

That there is a \textit{global} frame in which these conditions hold in a neighbourhood of the 
whole past light cone is a strong condition. It is possible that it can be weakened. 
For example, if a neighbourhood of the past light cone
can be covered by patches in each of which the 
field varies slowly in a null direction transverse to the light cone,
it might be possible to show that the contribution from each patch vanishes in the 
limit and thus to prove a more powerful result. 

\section{Curved Spacetime}

We assume again that the field $\phi$ is of compact support and we will
again split the region of integration $J^-(x)\cap \textrm{Support}(\phi)$ into three parts:
the deep chronological past, $W_3$, bounded away from $\del J^-(x)$;
$W_2$, a neighbourhood of $\del J^-(x)$ bounded away from $x$;
and the near region, $W_1$, a neighbourhood of $x$. 
We assume  certain differentiability 
and other conditions 
on $\phi$ and the metric which will be stated as they are used during the calculation. 

Let $N$ be a Riemann normal neighbourhood of $x$ with Riemann Normal Coordinates 
(RNC)  $\{y^\mu\}$ centred on the origin $x$ and, as before, we define
spatial polar coordinates: $r = \sqrt{\sum_{i= 1}^3 (y^i)^2}$, $\theta$ and $\varphi$.
We also again define radial null coordinates $u = \frac{1}{\sqrt{2}}(-y^0 - r)$
and $v =  \frac{1}{\sqrt{2}}(-y^0 + r)$ in $N$ where $u$ and $v$ increase 
into the past. 

We define $LC :=\del J^-(x)\cap \text{Support}(\phi)$
and assume that every point of $LC$ lies on a unique past directed null geodesic 
from $x$. This is a strong condition: generally there will be caustics on
$LC$.  
Each null geodesic generator of $LC$, $\gamma(\theta,\varphi)$,
is labelled by the polar angles $\theta$ and $\varphi$ and
has tangent vector, $T(\theta,\varphi)$, at $x$ with components in RNC:
 $T(\theta,\varphi)^\mu = \frac{1}{\sqrt{2}}(-1, \sin\theta\cos\varphi, \sin\theta\sin\varphi, \cos\theta)$.  
 The past pointing null tangent vectors at $x$ come in antipodal pairs, $( T(\theta,\varphi), T(\pi - \theta, \varphi + \pi))$, such that 
 $T(\theta,\varphi)^\mu \, T(\pi - \theta, \varphi + \pi)_{\mu} = - 1$.
From this we define Null Gaussian Normal Coordinates (NGNC) $\{V,U, \theta, \varphi\}$
\cite{Friedrich:1999aa} in a neighbourhood, $N_{LC}$, of $\del J^-(x)$ which contains
$LC$ and is bounded 
away from $x$. The coordinates $\theta$ and $\phi$ are the
labels of the null geodesic
generators of $\del J^-(x)$, $\gamma(\theta, \varphi)$.
The coordinate $V$ is the affine parameter along each $\gamma(\theta, \varphi)$
and is equal to $v$ (the RNC) along the generators in the overlap of the RNC and NGNC patches.
The transverse null coordinate $U$ is the affine parameter along
past pointing, ingoing null geodesics  from every point on $LC$ such that the tangent vector 
to the ingoing null geodesic 
at point $p$ on $\gamma(\theta,\varphi)$  is the vector $T(\pi-\theta, \varphi+\pi)$
at $x$, parallely transported to 
$p$ along $\gamma(\theta,\varphi)$.

In calculating the integral (\ref{Bbar}), we will need to know the behaviour of the
function $V(y)$, which is the volume of the causal interval between $x$ and $y$.\footnote{To
avoid confusion between $V(y)$ and NGNC coordinate $V$ we always write the 
volume function with its argument $y$.} For $y$ in the near region close to $x$, we can use the results of 
Myrheim and Gibbons and Solodukhin \cite{Myrheim:1978, Gibbons:2007aa} to expand
$V(y)$ in RNC.  For the region down the light cone, 
we show in Appendix \ref{dlcvolume}
that for $y$ in $N_{LC}$ with NGNC $\{V,U, \theta, \phi\}$
the limit of $U^{-2} V(y)$ as $U\rightarrow 0$ is finite and we 
denote $\lim_{U \rightarrow 0} U^{-2} V(y) = f_0(V, \theta, \varphi)$.
Indeed, if the causal interval between $x$ and $y$
is contained in a tubular neighbourhood of null geodesic $\gamma(\theta, \varphi)$ 
on which there are Null Fermi Normal Coordinates (NFNC) \cite{Blau:2006aa} then 
$V(y) = U^2 f_0(V, \theta, \varphi) + U^3 G(V, U, \theta, \varphi)$,
where $G$ is a continuous function. Furthermore, $f_0$ is an 
increasing function of $V$ and so therefore is
$V(y)$, for small enough $U$. 

Using this information we now define the regions $W_i$, $i = 1,2,3$.

Let the near region, $W_1$, be a subregion of $N$:
\begin{equation}
W_1 := \{y \in N  \,|\, 0\le u\le v\le a \} 
\end{equation}
for some $a>0$ such that $W_1$ is approximately flat, \textit{i.e.} the metric in 
RNC everywhere in $W_1$ is close to the Minkowski metric $\eta_{\mu\nu}$ in inertial coordinates. 

\begin{figure}[h]
\begin{center}
\includegraphics[scale=0.4]{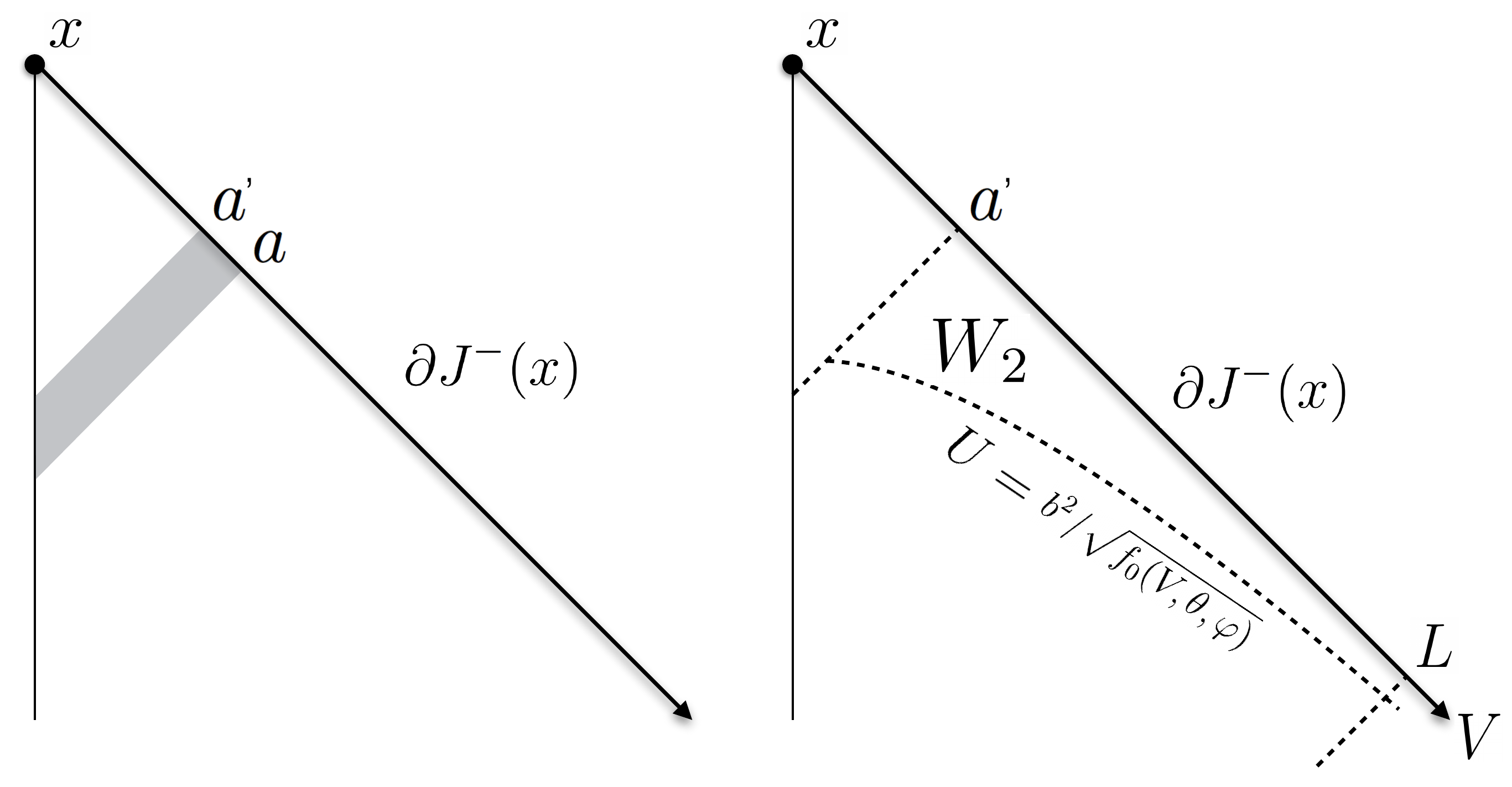}
\caption{The diagram on the left is a sketch of the overlap between regions $W_1$ and $W_2$, discussed in the text, as the shaded region. 
Note that here $a$ and $a'$ are the values of RNC and NFNC, $v$ and $V$, respectively. The diagram on the right shows the integration
region $W_2$.}
\label{curveregions}
\end{center}
\end{figure}

The down-the-light-cone region, $W_2$, is defined by 
\begin{equation}
W_2 := \{y \in N_{LC} \,|\, 0< a'(\theta, \varphi) \le V \le L,\quad 
\textrm{and}\quad 0\le U\le \frac{b^2}{\sqrt{f_0(V, \theta, \varphi)}} \} \,,
\end{equation}
where the cutoff $L$ is large enough that $W_2$ includes 
the whole of $LC$ outside $W_1$. The topology of $W_2$ is $I \times I \times S^2$, 
where $I$ is the unit interval.  $b>0$ is assumed to be 
small enough that the entire causal interval between 
the origin $x$ and any point with NGNC $(V,U, \theta, \varphi) \in W_2$ lies in a tubular neighbourhood of 
null geodesic, $\gamma(\theta, \varphi)$, on which 
Null Fermi Normal Coordinates (NFNC) exist. It is also assumed that $b$ is small enough that 
the correction  to $V(y)$ for $y \in W_2$ is small compared to the leading contribution, 
\textit{i.e.} $U^3 G(V, U, \theta,\varphi) \ll U^2 f_0(V, \theta, \varphi)$ in $W_2$. 
When the spacetime is flat, $u = U$ and $v =  V$ on the intersection of $N$ and $N_{LC}$
and taking $b = a' = a$ we recover the regions defined in the previous section for Minkowski space. 
When there is curvature,  $u\ne U$ and $v \ne V$ on the intersection of $N$ and $N_{LC}$
and there will be a mis-alignment between the boundaries of $W_1$ and $W_2$
for any choice of $a'$.
However, if the normal neighbourhood $N$ is approximately 
flat, then $u\approx U$ and $v \approx V$. The mismatch can be made
as small as we like by taking $a$ to zero as the density $\rho$ increases. 
We will keep $a'\ne a$, whilst bearing in mind that they will be almost equal.
This is sketched in Figure \ref{curveregions}

The deep chronological past, $W_3$, is 
\begin{equation}
W_3 := (\textrm{Supp}(\phi) \cap J^-(x) )\setminus (W_1\cup W_2)\,
\end{equation}
and is bounded away from $LC$. 

As before we define
\begin{align} I_i = \ohat
\int_{W_i}
                 & d^4y \;\sqrt{-g(y)}
                              \phi(y) e^{-\rho V(y)}\,,
\end{align}
for $ i = 1,2,3$ so that 
\begin{equation}
\bar{B}\phi(x):=
\fr{4\sqrt{\rho}}{\sqrt{6}}\big[-\phi(x)
+\rho(I_1 + I_2 + I_3)\big]\,.
\end{equation}

\subsection{Deep chronological past}\label{DCP}

Consider first 
\begin{align}
I_3 = \int_{W_3}
                 & d^4y\;
                              \sqrt{-g}\; \phi(y) e^{-\rho V(y)}\nn
&\times (1-9\rho V(y)+8(\rho V(y))^2-\frac{4}{3}(\rho V(y))^3)\,.
\end{align}
$V(y)$ is only zero on $LC$ and since $W_3$ is 
bounded away from $LC$, $V(y)$ is bounded away from zero on $W_3$: $V(y)\ge V_{min}> 0$. 
 So
%%%%%%
\begin{widetext}
\begin{align}
{}& \left\vert \int_{W_3}
                      d^4y\;
                              \sqrt{-g}\; \phi(y)\,
e^{-\rho V(y)} 
(1-9\rho V(y)+8(\rho V(y))^2-\frac{4}{3}(\rho V(y))^3)\right\vert \nn
 \le &\, e^{-\rho V_{\textrm{min}} }\int_{W_3} d^4y\;\sqrt{-g}
\left\vert \phi(y)
(1-9\rho V(y)+8(\rho V(y))^2-\frac{4}{3}(\rho V(y))^3) \right\vert 
\end{align}
\end{widetext}
which tends to zero faster than any power of $\rho^{-1}$ as $\rho \rightarrow \infty$. 

The conditions on $W_1$ and $W_2$ given above mean that, on $W_3$, $V(y)$ attains its
minimum on the  boundary between $W_3$ and $W_1\cap W_2$, and 
its approximate value is $V_{min} \approx \frac{\pi}{6} a^2 b^2$. 

\subsection{Down the light cone}
We work in NGNC, $\{V,U, \theta, \varphi\}$, in this region. 

We showed in Appendix \ref{dlcvolume} that in curved space
 the limit of $U^{-2} V(y)$ as $U\rightarrow 0$ is finite and 
we now assume enough differentiability of the metric so that $V(y)$
has the following expansion in $W_2$:
\begin{align}
V(y) &= U^2f_0(V,\theta,\varphi) + U^3 f_1(V,\theta,\varphi)+U^4 f_2(V,\theta,\varphi)+U^5 F(y)\,.\label{Volg}
\end{align}
We further assume enough differentiability of the metric and field 
that $\sqrt{-g(y)}$ and $\phi$ have the following expansions in $W_2$:
\begin{align}
\sqrt{-g(y)}&=h_0(V,\theta,\varphi)+Uh_1(V,\theta,\varphi)+U^2h_2(V,\theta,\varphi)
+U^3H(y)
\label{sqrtg}\\
\phi(y) &= \phi|_{U=0} + U\phi_{,U}|_{U=0}+
\fr{1}{2}U^2\phi_{,UU}|_{U=0}+ U^3\Phi(y)\,.
\label{phidlc}
\end{align}
The functions $F$, $H$ and $\Phi$ are continuous and we have
adopted a notation convention that 
a function denoted by a lower case letter is independent of $U$ and a 
function denoted by an upper case letter may depend on $U$. 

We will also use this expansion of the exponential factor in the integrand:
\begin{align}
e^{-\rho V(y)} = e^{-\rho U^2f_0}&\left(1-\rho U^3 \left(f_1+Uf_2+U^2F\right)+\fr{1}{2}\rho^2U^6\left(f_1+Uf_2+U^2F\right)^2\right.\nn
&\left.\;\;\;+\sum_{k=3}^\infty \fr{(-\rho)^k}{k!}U^{3k}\left(f_1+Uf_2+U^2F\right)^k\right).
\label{expV}
\end{align}

We want to calculate \begin{equation} \label{i2}
I_2 =  \ohat \int d\Omega_2\int_{a'}^L dV \int_0^{\frac{b^2}{\sqrt{f_0(V,\theta,\varphi)}}} 
dU  \sqrt{-g(y)} \phi(y) e^{-\rho V(y)}.
\end{equation}
Substituting the expansions (\ref{Volg}) - (\ref{expV}) into (\ref{i2}) one finds three types of integrals.
Integrals of the first kind, denoted by $I_{21}$, involve only $U$-independent unknown functions
 and do not have a factor of the infinite sum. A general such term can be written as
\be
I_{21}:= \ohat\left\{\rho^q \int d\Omega_2\int_{a'}^L dV\,  \Upsilon(V,\theta,\varphi) \int_0^{\frac{b^2}{\sqrt{f_0(V,\theta,\varphi)}}} 
dU \, U^{n+3q} e^{-\rho U^2f_0}\right\},
\ee 
where $\Upsilon$ denotes one (or a product) of the unknown functions independent of $U$, $q=0,1,2$ and $0\le n\le 4+q$. 
These terms can be dealt with straightforwardly since the $U$-integration can be done explicitly:
\begin{align}
I_{21}& =\ohat \left\{\fr{1}{2\rho^{\fr{n+q+1}{2}}} \int d\Omega_2
\int_{a'}^L dV\, \fr{\Upsilon(V,\theta,\varphi)}{f_0(V,\theta,\varphi)^{\fr{n+3q+1}{2}}}
\right.\nn
&\qquad\qquad\qquad\qquad\qquad\qquad\left.
\times\left(\Gamma\left(\fr{n+3q+1}{2}\right)
-\Gamma\left(\fr{n+3q+1}{2},\rho b^4\right)\right)\right\}.
\end{align}
The second term is exponentially small. The
first term is annihilated by $\ohat$ for $n+q=0,1,2$, and (after being multiplied by $\rho^{3/2}$) 
contributes a term that goes to zero in the limit  for $n+q>2$.

Integrals of the second kind, denoted by $I_{22}$, involve $U$-dependent unknown functions 
-- recall, these are denoted by capital letters -- 
and do not have a factor of the infinite sum in the integrand.
Each can be written generically as
\be
I_{22}:=  \ohat\left\{\rho^q  \int d\Omega_2\int_{a'}^L dV \int_0^{\frac{b^2}{\sqrt{f_0(V,\theta,\varphi)}}} 
dU \, U^{n+3q}\, \Xi(U,V,\theta,\varphi) e^{-\rho U^2f_0}\right\},
\ee
where $n\ge3$ when $q=0$, and $n\ge 2$ when $q=1,2$. 
Acting with $\ohat$ on $\rho^q e^{-\rho U^2 f_0}$ 
we find
\begin{align}
I_{22}&=\fr{\rho^{q}}{3} \int d\Omega_2\int_{a'}^L dV \int_0^{\frac{b^2}{\sqrt{f_0(V,\theta,\varphi)}}} 
dU \, U^{n+3q}\, \Xi(U,V,\theta,\varphi) e^{-\rho U^2f_0}\nn
&\qquad\times\left(3+11q+12q^2+4q^3-3\rho U^2 f_0 (3+2q)^2+12\rho^2U^4f_0^2(2+q)-4\rho^3U^6f_0^3\right)\,,
\end{align}
each term of which can be bounded. We show one example here:
\begin{align}
&\Big| \,\rho^{q} \int d\Omega_2\int_{a'}^L dV \int_0^{\frac{b^2}{\sqrt{f_0(V,\theta,\varphi)}}} 
dU \, U^{n+3q}\, \Xi(U,V,\theta,\varphi) e^{-\rho U^2f_0}\,\Big|\nn
&\le \rho^{q}  \int d\Omega_2
\int_{a'}^L dV\,  \| \Xi\|_U (V, \theta,\varphi) \,\int_0^{\frac{b^2}{\sqrt{f_0(V,\theta,\varphi)}}} 
dU \, U^{n+3q}\, e^{-\rho U^2f_0}\nn
&= \fr{1}{2\rho^{\fr{n+3q+1}{2}}} \int d\Omega_2
\int_{a'}^L dV\,
 \fr{ \| \Xi\|_U (V, \theta,\varphi) }{f_0(V,\theta,\varphi)^{\fr{n+3q+1}{2}}}\nn
 &\qquad\qquad\qquad\qquad\qquad\qquad\times
\left(\Gamma\left(\fr{n+3q+1}{2}\right)-\Gamma\left(\fr{n+3q+1}{2},\rho b^4\right)\right), \label{fzero}
\end{align}
where $\| \Xi\|_U(V, \theta, \varphi)$ is the uniform norm of $\Xi$ over the $U$ 
coordinate.  After being multiplied by $\rho^{3/2}$, this goes to zero in the limit, 
since $n+3q\ge 3$. The other terms are similar. 

Finally, the remaining term in $I_2$ is 
\begin{align}
I_{23}:= &\ohat  \int d\Omega_2\int_{a'}^L dV \int_0^{\frac{b^2}{\sqrt{f_0(V,\theta,\varphi)}}} 
dU \, \sqrt{-g(y)} \phi(y) e^{-\rho U^2f_0}\nn
&\qquad\qquad\qquad\qquad\qquad\times\sum_{k=3}^\infty \fr{(-\rho)^k}{k!}U^{3k}G(y)^k.
\end{align}
where $G(y) = 
\left(f_1+Uf_2+U^2F\right)$.

We will see that each term in $I_{23}$ arising from the action of $\ohat = (1 + 9H_1
+ 8 H_2 + \frac{4}{3}H_3)$ on the integrand, 
is $O(\rho^{-2})$.
First we note that
\be
\big|\sum_{k=3}^\infty \fr{(-\rho)^k}{k!}U^{3k}{G}^k\big|\le \frac{\rho^3}{6}U^9
|G|^3e^{\rho U^3|G|}\,,
\ee
where $G: = f_1+Uf_2+U^2F$.
Recall that in defining $W_2$, $b$ was chosen small enough that 
 $U^3|G| \ll U^2 f_0$ in $W_2$, so we  have
 \begin{align}
&\Bigg|\int d\Omega_2\int_{a'}^L dV \int_0^{\frac{b^2}{\sqrt{f_0(V,\theta,\varphi)}}} 
dU \, \sqrt{-g(y)} \phi(y) e^{-\rho U^2f_0}\nn
&\qquad\qquad\qquad\qquad\qquad\times\sum_{k=3}^\infty \fr{(-\rho)^k}{k!}U^{3k}G^k\Bigg|\nn
&\le 
\frac{ \rho^3}{6} \int d\Omega_2  \int_{a'}^L dV  \int_0^{\frac{b^2}{\sqrt{f_0}}} dU  
\,\left\vert \sqrt{-g(y)} \phi(y)G^3\right\vert U^9e^{-\frac{\rho}{2} U^2 f_0}  \nn
& \le \frac{ \rho^3}{6} \,\rho^{3} \int d\Omega_2 \int_{a'}^L dV\,
 \Vert\sqrt{-g}\,\phi G^3\,\Vert_U \int_0^{\frac{b^2}{\sqrt{f_0}}} dU \,
 U^9 e^{-\frac{\rho}{2} U^2 f_0}\nn
 & = \fr{2^6}{\rho^2}  \int d\Omega_2 \int_{a'}^L dV\,\fr{\Vert\sqrt{-g}\,\phi G^3\,\Vert_U}{f_0(V,\theta,\varphi)^5}
\end{align}
neglecting terms proportional to $e^{-\rho b^4/2}$.
Exchanging the order of 
summation and integration is justified as the range of integration is finite and the partial sums of the series are uniformly integrable.

After being multiplied by $\rho^{3/2}$ this term 
is of order $\rho^{-1/2}$ and hence goes to zero in the limit. The terms 
arising from the action of each $H_i$, $i=1,2,3$ on the integrand can be
 bounded similarly and are also of order $\rho^{-1/2}$.

\subsection{The near region}

Now that it has been demonstrated that the contributions to the 
mean from the region of integration bounded away from the origin
vanish in the limit, 
we can conclude that the value of $\lim_{\rho\rightarrow \infty} \bar{B}\phi(x)$, 
if it is finite, must be local 
since as $\rho \rightarrow \infty$, we can choose $a$ to be arbitrarily small. 
The only local scalar quantities of the correct dimensions are $\Box\phi(x)$
and $R\phi(x)$.
In this section we show that the limit is finite and that the precise linear
combination is (\ref{lim}).

In the near region, $W_1$, we work with
Riemann normal coordinates $\{y^\mu\}$ centred on $x=0$ and the usual
spatial polar coordinates: $r = \sqrt{\sum_{i= 1}^3 (y^i)^2}$, $\theta$ and $\varphi$, and 
 radial null coordinates $u = \frac{1}{\sqrt{2}}(-y^0 - r)$
and $v =  \frac{1}{\sqrt{2}}(-y^0 + r)$.
We will show that 
\be \lim_{\rho \rightarrow \infty} \left(\rho^{3/2} I_1 - \rho^{1/2} \phi(x)\right) 
=\frac{\sqrt{6}}{4}\left( \Box\phi(x) - \frac{1}{2}R(x)\phi(x)\right),
\ee
where
\begin{align}
I_1 &= \ohat \int_{W_1} d^4 y \sqrt{-g(y)}\phi(y) e^{-\rho V(y)} \nn
&= \ohat \int_{0}^a dv \int_{0}^v du \int d\Omega_2
\fr{(v-u)^2}{2} \sqrt{-g(y)} \phi(y) e^{-\rho V(y)} \,.\label{eq5.1}
\end{align}
We note that above and throughout this subsection $\sqrt{-g(y)}$ will  denote the square root of minus the determinant of the metric \textit{in RNC}.

In $W_1$, we have  expansions  \cite{Myrheim:1978, Gibbons:2007aa}:
\begin{align}
\sqrt{-g} &= 1 - \fr{1}{6}y^{\mu}y^{\nu}R_{\mu\nu}(0)+y^{\mu}y^{\nu}y^{\rho}T_{\mu\nu\rho}(y).
\label{eq5.2}\\
\phi(y) &= \phi(0) + y^{\mu}\phi_{,\mu}(0)+
\fr{1}{2}y^{\mu}y^{\nu}\phi_{,\mu\nu}(0)\nn 
&\quad+ y^{\mu}y^{\nu}y^{\alpha}\Psi_{\mu\nu\alpha}(y)
\label{eq5.3}\\ 
 V(y) &= \frac{\pi}{24}\tau^4 - \fr{\pi}{4320}\tau^6R(0)
+\fr{\pi}{720}\tau^4y^{\mu}y^{\nu}R_{\mu\nu}(0)\nn
&\quad+\tau^4y^{\mu}y^{\nu}y^{\rho}
S_{\mu\nu\rho}(y)
=V_0(y)+\delta V(y)
 \label{eq5.4}
\end{align}
where $\tau^2 = 2 u^2 v^2$,  $V_0(y)=\frac{\pi}{24}\tau^4= \frac{\pi}{6} u^2 v^2$ and $\delta V(y)$ is the rest. 
$T_{\mu\nu\rho}(y)$, $\Psi_{\mu\nu\alpha}(y)$ and $S_{\mu\nu\rho}(y)$ 
are $C^3$-functions.

We also use the expansion of the exponential factor,
\be
e^{-\rho V(y)}=e^{-\rho V_0(y)}e^{-\rho\delta V(y)}
=e^{-\rho V_0(y)}\big(1-\rho\delta V(y) %+\fr{1}{2}\rho^2\delta V(y)^2
+ \sum_{k=2}^{\infty}\fr{(-\rho)^k}{k!}(\delta V)^k\big).
\label{eq5.5}
\ee
%\bwt
Using (\ref{eq5.2})-(\ref{eq5.5}) we expand the integrand in 
(\ref{eq5.1}) and collect the terms in 4 groups:
\be
\sqrt{-g(y)} \phi(y) e^{-\rho V} = \left(A(y) + B(y) + C(y) + D(y) \right) e^{-\rho V_0}\,,
\ee 
where
\begin{align}
A(y) = &\, \phi+\fr{1}{2}y^{\mu}y^{\nu}\phi_{,\mu\nu}-\fr{1}{6}\phi\, y^{\mu}y^{\nu}R_{\mu\nu}
+\fr{\rho\pi\tau^4}{4320}\phi(\tau^2R-6y^{\mu}y^{\nu}R_{\mu\nu})\,;\\
B(y) =& \, \left(1 - \frac{1}{6} y^\mu y^\nu R_{\mu\nu}\right) y^\alpha \phi_{,\alpha}
- \frac{1}{12} y^\mu y^\nu y^\alpha y^\beta R_{\mu\nu}\phi_{, \alpha\beta} \nn
 &\!\!\!\!\!\!\!\!\!\!\!\!\!\!\!\!\!\!+\rho
\left(y^\alpha\phi_{,\alpha}+\fr{1}{2}y^{\alpha}y^{\beta}\phi_{,\alpha\beta}-\fr{1}{6}\phi\, y^{\mu}y^{\nu}R_{\mu\nu}
-\fr{1}{6}\, y^{\mu}y^{\nu}y^\alpha R_{\mu\nu}\phi_{,\alpha}
-\fr{1}{12}\, y^{\mu}y^{\nu}y^{\alpha}y^{\beta}R_{\mu\nu}\phi_{,\alpha\beta}
\right)\nn
&\times\left(\fr{\pi}{4320}\tau^6R-\fr{\pi}{720}\tau^4y^\rho y^\sigma R_{\rho\sigma}\right)\,;
\end{align}
\begin{align}
C(y) =
& \left(1-\frac{1}{6}y^{\mu}y^{\nu}R_{\mu\nu}+y^{\mu}y^{\nu}y^{\alpha}T_{\mu\nu\alpha}\right)y^{\rho}y^{\delta}y^{\sigma}\Psi_{\rho\delta\sigma}\nn
&\!\!\!\!\!\!\!\!\!\!\!\!\!\!\!\!\!\!
+y^{\mu}y^{\nu}y^{\alpha}T_{\mu\nu\alpha}\left(\phi(0)+y^{\mu}\phi_{,\mu}(0)+\frac{1}{2}y^{\mu}y^{\nu}\phi_{,\mu\nu}(0)\right) \nn
&\!\!\!\!\!\!\!\!\!\!\!\!\!\!\!\!\!\!-\rho\left\{\left(y^{\mu}y^{\nu}y^{\alpha}\Psi_{\mu\nu\alpha}-\frac{1}{6}y^{\eta}y^{\sigma}R_{\eta\sigma}y^{\mu}y^{\nu}y^{\alpha}\Psi_{\mu\nu\alpha}+\phi y^{\mu}y^{\nu}y^{\alpha}T_{\mu\nu\alpha}\right)\left(-\frac{\pi}{4320}\tau^{6}R+\frac{\pi}{720}\tau^{4}y^{\mu}y^{\nu}R_{\mu\nu}\right)\right.\nn
&\!\!\!\!\!\!\!\!\!\!\!\!\!\!\!\!\!\!\left.+\left(1-\frac{1}{6}y^{\eta}y^{\sigma}R_{\eta\sigma}\right)\tau^{4}y^{\alpha}y^{\beta}y^{\gamma}S_{\alpha\beta\gamma}\left(\phi+y^{\mu}\phi_{,\mu}(0)+\frac{1}{2}y^{\mu}y^{\nu}\phi_{,\mu\nu}(0)+y^{\alpha}y^{\beta}y^{\gamma}\Psi_{\alpha\beta\gamma}\right)\right.\nn
&\!\!\!\!\!\!\!\!\!\!\!\!\!\!\!\!\!\!\left.+\phi y^{\mu}y^{\nu}y^{\alpha}T_{\mu\nu\alpha}\tau^{4}y^{\alpha}y^{\beta}y^{\gamma}S_{\alpha\beta\gamma}+y^{\mu}y^{\nu}y^{\alpha}T_{\mu\nu\alpha}\left(y^{\mu}\phi_{,\mu}(0)+\frac{1}{2}y^{\mu}y^{\nu}\phi_{,\mu\nu}(0)+y^{\mu}y^{\nu}y^{\alpha}\Psi_{\mu\nu\alpha}\right)
\right.\nn
&\left.\times\left(-\frac{\pi}{4320}\tau^{6}R+\frac{\pi}{720}\tau^{4}y^{\mu}y^{\nu}R_{\mu\nu}+\tau^{4}y^{\alpha}y^{\beta}y^{\gamma}S_{\alpha\beta\gamma}\right)\right\}\,;\\
D(y) =&\, \sqrt{- g(y)} \phi(y) \sum_{k=2}^{\infty}\fr{(-\rho)^k}{k!}(\delta V)^k\,.
\end{align}
%\ewt
Then,
\be
I_1 =I_A + I_B + I_C + I_D
\ee
where
\begin{align}
I_A := \ohat \int_{0}^a dv \int_{0}^v du  \int d\Omega_2 \frac{(v-u)^2}{2}  A(y)e^{-\rho V_0}\,,
\end{align}
and similarly for $I_B$, $I_C$ and $I_D$. 
$I_A$ and $I_B$ are doable integrals involving no unknown 
functions. We will see that $\rho^{3/2} I_A$  gives the nonzero contributions in the
limit and $\rho^{3/2} I_B$,   $\rho^{3/2} I_C$ and $\rho^{3/2} I_D$
vanish in the limit.
\bwt
Consider $I_A$. 
Integrating over the angular coordinates gives 
\begin{align}
I_A&=\ohat\int_{0}^a dv \int_{0}^v du \fr{(v-u)^2}{2}
\left(4\pi\phi+\pi(u+v)^2\big(\phi_{,00}-\fr{1}{3}(1+\fr{\pi}{30}\rho\,u^2v^2)\phi\,R_{00}\big)\right.\nn
&\left.+\fr{\pi}{3}(v-u)^2\big(\phi_{,ii}-\fr{1}{3}(1+\fr{\pi}{30}\rho\,u^2v^2)\phi\, R_{ii}\big)+\fr{\pi^2}{135}\rho\, u^3v^3\phi\, R
\right)e^{-\rho\fr{\pi}{6} u^2v^2}.
\end{align}
\ewt
Expanding out the brackets in the integrand gives a sum of terms each of which is a constant
times something of the form
\begin{align}
&\ohat\int_{0}^a dv \int_{0}^v du \;u^mv^n (\sigma u^2 v^2)^q e^{-\sigma u^2v^2}\\
= \, & \, (-1)^q H^q \ohat  \int_{0}^a dv \int_{0}^v du \;u^mv^n  e^{-\sigma u^2v^2}\,,\label{Hohat}
\end{align}
where $\sigma = \rho \pi/6$, $q = 0, 1$ and $m+n = 2, 4$. $H$ commutes with $\ohat$ and does not change the power of $\sigma$ so we only need look at case $q=0$. 

Let 
\be\label{Imn}
Z_{m,n}:=\int_{0}^a dv \int_{0}^v du \;u^mv^n e^{-\sigma u^2v^2},\quad m,n\in \mathbb{N}\,.
\ee
 Up to exponentially small terms we have, for $n\ne m$,
 \begin{align}
\label{inm1}
Z_{m,n} =\fr{1}{2(m-n)}\left[
\fr{1}{\sigma^{\fr{m+n+2}{4}}}\Gamma\big(\fr{m+n+2}{4}\big)
-\fr{a^{n-m}}{{\sigma}^{\fr{m+1}{2}}}
\Gamma\big(\fr{m+1}{2}\big)
\right]
\end{align}
and, for $n=m$,
\be \label{Inn}
Z_{m,m} = 
\frac{1}{8\sigma^{\fr{m+1}{2}}}\,\left(\ln(\sigma\, a^4)-\Psi\left(\fr{m+1}{2}\right)\right)\Gamma\left(\fr{m+1}{2}\right)\,.
\ee

The terms in $Z_{m,n}$ with $n\ne m$ 
are powers of $\sigma$ and since $\ohat$ kills $\sigma^{-\frac{1}{2}},
\sigma^{-1}, \sigma^{-\frac{3}{2}}$, 
the correction to  the limit  of $\rho^{\frac{3}{2}}I_B$ from such terms is
$O(\rho^{ -\frac{1}{2} })$.
The contributions that are nonzero in the limit come from terms proportional to 
$Z_{m,m}$ for $m = 1$, $2$  and arise from the action of $\ohat$ on 
 $\log(\sigma)/\sigma^{\frac{m+1}{2}}$. 
We find that 
\be
\lim_{\rho \rightarrow \infty} \fr{4}{\sqrt{6}}\left(\rho^{3/2} I_A-\sqrt{\rho}\phi(0)\right)
=(\Box-\fr{1}{2}R(0))\phi(0).
\ee

Consider $I_B$. Again there are no unknown functions, the integration over the angles can be 
done and 
$I_B$ becomes a sum of terms, each of which is a constant times something
of the form (\ref{Hohat}) where $q = 0,1$ and $m+n = 3, 5, 6,7,8$.
 Using (\ref{inm1}) and (\ref{Inn}) we find that the leading contribution
to $\rho^{3/2} I_B$ is a sum of terms that are $O(\log\rho/\sqrt{\rho})$
which vanishes in the limit. 
%{\bf DB: Fay, I can give more details here if you think it would be useful.}

%and integrals over W_1, with integrand 
%u^i * v^j exp(- \sigma u^2v^2), go like (see equation 89)
%
%(1) 1/(i-j) * ( 1/\rho^{(i+j+2)/4} + a^{i-j}/\rho^{(i+1)/2} )
%
%for i?j. Now the second term can never give \rho^{-1/4} since for i=0,1,2 Ohat kills it, and for i>2 it
%goes at least like \rho^{-2}. So we only need to show that the first term never contributes (at least when 
%i+j=5, but in fact it never does). 
%To show this note that for each term in the expansion of t^n * r^2m+2 of the form u^i*v^j, there will be another term 
%with the same coefficient (because of the symmetry u<->v) of the form u^j * v^i. But because of the
%over factor of (i-j)^-1 in (1) these two terms will sum to zero. 
%
%The only exception occurs when n+2m+2 is even so that there is a ``middle? term that goes like u^k*v^k and is 
%not paired with any other term. In this case let 2k:=n+2m+2, then
%one can see that this case appears in the corrections only when k?3 and therefore the corresponding integral over W1 
%gives (see equation 90)

%Again it is a linear
%combination of integrals $I_{m,n}$ of which only 
%those for which $n=m$ are not exponentially suppressed,
%and from (\ref{Inn}) we see that 
%$I_{m,m} = O(\rho^{-\frac{m+1}{2}} \log\rho)$ in the limit. 
%It is straightforward then to show that 
%\be
%\rho^{3/2}\ohat I_B= O(\fr{\log\rho}{\sqrt{\rho}})
%\ee
%as $\rho\rightarrow \infty$. 
 
Consider $I_C$. It is a sum of terms, each of the form
\begin{align}
& \ohat\left\{\int_{0}^a dv \int_{0}^v du \int d\Omega_2
\fr{(v-u)^2}{2} (y^\alpha)^s(\sigma u^2v^2)^q\, \Xi(y) e^{-\sigma u^2v^2}\right\}\\
= &\, (-1)^q H^q \ohat\left\{\int_{0}^a dv \int_{0}^v du \int d\Omega_2
\fr{(v-u)^2}{2} (y^\alpha)^s\, \Xi(y) e^{-\sigma u^2v^2}\right\},
\end{align}
where $q = 0, 1$, $s = 3,\dots9$, $(y^\alpha)^s$ stands for a product of $s$ coordinates and
 $\Xi$ stands for one of the functions of  $y$ in the expansion of the brackets in term $C(y)$. 
 We have used  $\tau^4=4u^2v^2$ and 
 $\sigma = \rho \pi/6$.

The operator $H$ does not change any power of $\sigma$ 
and we see that what it acts on is the same as (\ref{nearint}), except that 
in that earlier case $s=3$ only. 
Thus we need to calculate terms like (\ref{nearcorrK}) with $m+n = s= 3,\dots 9$. 
Using the calculations in subsection \ref{flatnear} we find that $I_C$ vanishes in the limit
and the leading corrections are of order  $\rho^{-1/4}$.

\bwt
Finally we deal with $I_D$,
\begin{align}
I_D&=\ohat\int_0^adv\int_0^v du\int d\Omega_2\, (v-u)^2 f(y)\sum_{k=2}^{\infty}\fr{(-\rho\delta V)^k}{k!}
e^{-\rho \frac{\pi}{6} u^2 v^2}
%\\
%&=\int_0^adu\int_0^adv \int d\Omega_2f(y)(-1+\rho\tau^4\bar{S}(y)-\rho^2\tau^8\bar{S}(y)^2
%+e^{\rho\tau^4\bar{S}(y)})
%e^{-\rho V_0}
\end{align}
\ewt
where $f(y):=\sqrt{-g(y)}\phi(y)/2$. We 
use $\sigma = \rho \pi/6$, as before,  and define $\xi(y)$ and $\Lambda_{\mu\nu}(y)$ by 
\begin{align}
\rho \delta V(y)& = \sigma u^2 v^2 \xi(y)\\
\xi(y) & = y^\mu y^\nu \Lambda_{\mu\nu}(y)\,.
\end{align}
We have assumed that in $W_1$ the metric is 
approximately flat so that $\delta V \ll V_0$ and $\|\xi\|_1\ll 1$ in $W_1$. This 
will be important later.  
%{\color{red}{\bf(DB: do we need to sort out notation $(v-u)^2\sqrt{-g}$)}?} 
Now, leaving the angular integration to be done at the end, we have 
\begin{align}
%I_{D}	&=	
&\hat{\mathcal{\mathcal{O}}}\int_0^a dv\int_0^v du\sum_{k=2}^{\infty}\frac{1}{k!}
(v-u)^2 f(y)\xi^{k}(-\sigma u^2 v^2)^k e^{-\sigma u^{2}v^{2}} \nn
&=	\hat{\mathcal{\mathcal{O}}}\int_0^a dv\int_0^v du\sum_{k=2}^{\infty}\frac{1}{k!}
(v-u)^2 f(y)\xi^{k}H_{k}e^{-\sigma u^{2}v^{2}}\nn
&=\sum_{k=2}^{\infty} \frac{1}{k!} H_{k} \ohat \int_0^a dv\int_0^v du\,
(v-u)^2 f(y)\xi^{k} e^{-\sigma u^{2}v^{2}}\nn
&=-\frac{1}{6}\int_0^a dv\int_0^v du\,\sum_{k=2}^{\infty} \frac{1}{k!}H_{k}\left\{u^{3}\frac{\partial^{3}}{\partial u^{3}}
\left((v-u)^2 f(y)\xi^{k}\right)\,e^{-\sigma u^{2}v^{2}}\right\}\nn
&\qquad\qquad\qquad\qquad+\frac{1}{3}\int_0^a dv\,\sum_{k=2}^{\infty} \frac{1}{k!} H_{k}\left\{ v^{6} \tilde{f}(2v^2\tilde{\Lambda}_{tt})^{k}e^{-\sigma v^{4}}\right\},\label{IDbound}
\end{align}
where we have used that $\ohat$ and $H_k$ commute and
for the last step we used the same integration by parts done in equations (\ref{nearcorr})-(\ref{Kmn}). The tilde means setting $u=v$, so that $\tilde{\xi} =2v^2\tilde{\Lambda}_{tt}$.
The exchanging of the orders of summation, integration and differentiation by $\sigma$ is 
justified as the partial sums are uniformly integrable.

Let us now consider the first term in (\ref{IDbound}). Acting with the $H_{k}$ we find it equals 
\begin{align}
&-\frac{1}{6}\int_0^a dv\int_0^v du\,\sum_{k=2}^{\infty}\frac{(-\sigma)^{k}}{k!}u^{2k} v^{2k} u^{3}\frac{\partial^{3}}{\partial u^{3}}
((v-u)^2 f(y)\xi^{k})\,e^{-\sigma u^{2}v^{2}}\nn
&= - \fr{\sigma^2}{6}\int_0^a dv\int_0^v du\, u^7 v^4 \sum_{k=0}^{\infty}\frac{(-\sigma)^k}{(k+2)!}u^{2k} v^{2k}\frac{\partial^{3}}{\partial u^{3}}
[(v-u)^2 f(y)\xi^{k+2}]\,e^{-\sigma u^{2}v^{2}}
\label{id1}
\end{align}
%where $g(y):=(v-u)^2f(y)$. 
Our  strategy  will be to bound the infinite sum by something times 
the exponential $e^{\sigma u^2 v^2/2} $, which is possible because $|\xi|\ll 1$. Then the 
resulting Gaussian integral can be done. 

To this end we expand 
\begin{align}
\fr{\del^3}{\del u^3}(h\,\xi^{k+2})&=\left[h'''\xi^{2}+3(k+2)h''\xi\xi' + 3(k+2)(k+1)h'(\xi')^2 
+ 3(k+2)h'\xi\xi''\right. \nn
&\left.+ 3(k+2)(k+1)h\xi'\xi''  +  (k+2)h\xi\xi'''\right]\xi^k + k(k+2)(k+1)h(\xi')^3 \xi^{k-1},
\label{du3}
\end{align}
where $h = (v-u)^2 f$ and $'$ denotes differentiation with respect to $u$.
Recalling that $\xi(y)=y^\mu y^\nu\chi_{\mu\nu}(y)$, we see that 
each of the terms multiplying $\xi^k$ has a factor $u^iv^j$
with $3\le i+j\le 6$ and each 
term multiplying $\xi^{k-1}$ has a factor of $u^iv^j$, $5\le i+j\le 8$.  

Thus, every term in (\ref{id1}) is of the 
form 
\begin{align}
{\sigma^{2+q}}\int_0^a dv\int_0^v du\, \Theta(y) u^{7 + 2q+i} v^{4+ 2q+j} \sum_{k=0}^{\infty}\frac{(-\sigma)^k}{(k+2+q)!}u^{2k} v^{2k} \xi^k 
\,e^{-\sigma u^{2}v^{2}}
\label{idbound1}
\end{align}
where  $3\le i+j\le6$ when $q = 0$  and $5\le i+j\le 8$ when $q = 1$.

As $|\xi| \ll 1$ in $W_1$ it is certainly less than $1/2$ and we can bound
the infinite sum 
\be\label{bou2}
\Big|\sum_{k=0}^{\infty}\frac{\xi^{k}}{(k+2+ q)!}(-\sigma)^{k}u^{2k} v^{2k}\Big| 
< e^{\frac{1}{2} \sigma u^2 u^2 } 
\ee
$\forall y\in W_1$. 
Then (\ref{idbound1}) for $q=0$  can be bounded:
\begin{align}
&\Big|{\sigma^{2}}\int_0^a dv\int_0^v du\, \Theta(y) u^{7+i} v^{4+j} \sum_{k=0}^{\infty}\frac{(-\sigma)^k}{(k+2)!}u^{2k} v^{2k} \xi^k
\,e^{-\sigma u^{2}v^{2}}\Big| \\
& < \|\Theta \|_1 \sigma^{2} 
\int_0^a dv\int_0^v du\,  u^{7+i} v^{4+j}  
\,e^{-\frac{1}{2}\sigma u^{2}v^{2}}\nn
&=\left\{
 \begin{array}{l}
\fr{2\|\Theta\|_1}{(i-j+3)}\left(\fr{\Gamma((13+i+j)/4)}{\hat\sigma^{(5+i+j)/4}}-\fr{a^{j-i-3}\Gamma(4+i/2)}{\hat\sigma^{2+i/2}}\right)
,\;\quad\quad \;\,i-j+3\ne 0\\
\frac{\|\Theta\|_1}{\hat\sigma^{2+i/2}}\,\left(\ln(\hat\sigma\, a^4)-\Psi\left(4+i/2\right)\right)
\Gamma\left(4+i/2\right)
,\,\;\; i-j+3= 0\\
 \end{array} \right\}\;,
\label{Iij}
\end{align}
up to exponentially small terms, where $\hat\sigma = \sigma/2$. The integral over the 
angles contributes a factor of $4\pi$. 

Since $i+j \ge 3$, these terms, after multiplication by $\rho^{3/2}$, vanish
in the limit and the leading correction to the limit is $O(\ln(\rho)/\sqrt{\rho})$.
The terms for $q=1$ also tend to zero and 
the leading correction is again  $O(\ln(\rho)/\sqrt{\rho})$.

Let us now turn to the second term in \eqref{IDbound}:\\
\begin{align}
& \Big|\int_0^a dv\,\sum_{k=2}^{\infty}\frac{1}{k!}H_{k}\left\{ v^{6} \tilde{f}(2v^2\tilde{\Lambda}_{tt})^{k}e^{-\sigma v^{4}}\right\}\Big| \nn
&\le\int_0^a dv\, v^6  | \tilde{f} |
\sum_{k=2}^{\infty}\frac{1}{k!} 2^k v^{2k}|\tilde{\Lambda}_{tt}|^{k}\sigma^{k}v^{4k} e^{-\sigma v^{4}}\nn 
&\leq 4\sigma^{2} \,\|\tilde{f}\|_{1} \,||\tilde{\Lambda}_{tt}^{2}||_1\int_0^a dv\,v^{6}v^{12}\,e^{-\sigma v^{4}} 
\sum_{k=0}^{\infty}\fr{(2v^{2}|\tilde{\Lambda}_{tt}|)^{k}}{(k+2)!}\sigma^{k}v^{4k} \nn
&\leq 4\sigma^{2} \,\|\tilde{f}\|_{1} \,||\tilde{\Lambda}_{tt}^{2}||_1\int_0^a dv\,v^{6}v^{12}\,e^{-\sigma v^{4}} 
 e^{\frac{1}{2}\sigma v^4} \nn
& \leq 4\sigma^{2}\, \|\tilde{f}\|_{1} \,||\tilde{\Lambda}_{tt}^{2}||_1\int_0^a dv\,v^{18} e^{-\frac{1}{2}\sigma v^{4}}\,,\label{imijbndry}
\end{align}
where the second to last inequality follows from $2v^2 |\tilde{\Lambda}_{tt}| \ll 1$.
The final integral gives a contribution of order  $\rho^{-5/4}$, after multiplication by $\rho^{3/2}$
and is not one of the leading order corrections.

\subsection{Finite $\rho$ corrections}\label{finiterho}

As in the flat spacetime case, we want to know  
how $\bar{B}\phi(x)$ behaves when $\rho = l^{-4}$ is large but finite.
Unlike in flat spacetime however, since we lack an explicit expression for the expansion
of the volume of long skinny intervals ``down the light cone" in $W_2$,
the finite $\rho$ corrections to the limit 
from $W_2$ can only be given in terms of integrals of 
unknown functions, such as $f_0(V, \theta, \phi)$ in 
(\ref{fzero}), and are not very enlightening. In Appendix \ref{dlcvolume} it is 
shown that 
$f_0(V, \theta, \phi)$ depends only on the curvature components along
the null geodesic on the light cone labelled by $(\theta, \phi)$ and it 
may be possible to find $f_0(V, \theta,\phi)$ and further terms in the
expansion of the volume as expressions 
involving integrals of these curvature components along the null geodesic.

There is one case in which we do not need to know the
behaviour of the volume of long skinny intervals to bound
the corrections and that is when  
a neighbourhood of the whole of the light cone is covered by RNC in which 
the metric is approximately flat. That means that all corrections to the metric in RNC 
are small: 
$R_{\mu\nu }(0)y^\mu y^\nu \le 1$ and $\nabla_\alpha R_{\mu\nu} (0) y^\mu y^\nu y^\alpha \ll 1$
\textit{etc.} throughout $W_2$.  In particular, ${\cal{R}} L^2 \ll 1$ where ${\cal{R}}$ stands for any component of
the curvature at the origin and $L$ is the cutoff on the coordinate $v$ in $W_2$. 

Then, the NGNC coordinates $U$ and $V$ are replaced by their RNC versions 
$u$ and $v$ in the integrals for the corrections
and the expansions of all functions are well approximated by their flat space versions. For example,
 function $f_0(V^2, \theta, \phi) $ is well-approximated by 
$\frac{\pi}{6} v^2$. The integrals for the corrections in region $W_2$ 
can be bounded and the corrections take the same form as they do in the flat space case, 
up to factors of order one. 
In this case, however, the condition that the metric be approximately flat in RNC in the 
whole of $W_2$  means that the Ricci scalar times $\phi$, $R \phi $, at the origin is 
negligible compared with $\Box \phi$ -- assuming that the scale on which $\phi$ varies is small compared 
to the cutoff -- and the limiting value in this case is close to $\Box \phi$, as one would 
expect. 

Then,  the same conditions from the flat space apply, namely that $\phi$ 
should be slowly varying 
in the direction transverse to the light cone over the scale $a$ and $a \gg l$. And the same 
conditions as in the flat space case will also result from considering corrections 
from the near region, $W_1$, as
one can verify by examining the various terms. The conclusion is that
the situation in which we can estimate the corrections -- without further knowledge about long 
skinny intervals -- is when the region is approximately flat in some frame, and the field is 
slowly varying on the discreteness scale {in that same frame}.  In which case, the 
result is close to the flat space result and $\bar{B}\phi$ is close to $\Box \phi$. 

Consider now the case when $\phi = 1$. Then  the limit of $\bar{B}\phi$ equals $-\frac{1}{2} R$ and 
we can ask when the value of $\bar{B}\phi$ is close to that limit. Again, our lack of 
knowledge about $V(y)$ for long skinny intervals means we can only answer the question 
under conditions that the metric is approximately flat in RNC in the 
whole of $W_2$. In that case, if all curvature length scales are large compared with the discreteness
scale, then the value of $\bar{B}\phi$ is close to $-\frac{1}{2} R$. 

\section{Discussion}

In \cite{Dowker:2013vl,Glaser:2013sf}, causal set d'Alembertians 
were defined for dimensions $d=3$ and $d>4$, and it was shown that if the mean of
these d'Alembertians has a local limit as $\rho \rightarrow \infty$ then 
that limit will be $\Box - R/2$ in all dimensions. 
We expect the argument for the existence of the local 
limit under certain conditions given above for $d=4$, to be able to be generalised for 
$d=3$ and $d>4$.  
In two dimensions, the conformal flatness of spacetime should
make the proof more straightforward. 

So far we have ignored the important question of the fluctuations about the mean. 
These fluctuations grow with the sprinkling density.
In order to tame the fluctuations, Sorkin introduced, in two dimensions, for each fixed
physical non-locality length scale, $l_k\ge l$, a causal set operator,
$B^{(2)}_k$ \cite{Sorkin:2007qi}, whose mean over sprinklings at density $\rho$ does not 
depend on $\rho$ but is equal to the 
mean of $B^{(2)}$ with $\rho$ replaced by $l_k^{-2}$.
This was extended to four dimensions  \cite{Benincasa:2010ac}:
 \be\label{nonlocalB}
B^{(4)}_{k}\phi(x)=\frac{4}{\sqrt{6}l_k^2}\bigg[-\phi(x)+
\epsilon\sum_{y\prec x}f(n(x,y),\epsilon)\phi(y)\bigg],
\ee
where $\epsilon=(l/l_k)^4$ and
\begin{equation}
f(n,\epsilon)=(1-\epsilon)^n 
\left[ 
1-\frac{9\epsilon n}{1-\epsilon}
+\frac{8\epsilon^2n!}{(n-2)!(1-\epsilon)^2}
-\frac{4\epsilon^3n!}{3(n-3)!(1-\epsilon)^3}
\right]
\,,
\end{equation}
and also to all 
 other dimensions \cite{Dowker:2013vl,Glaser:2013sf}. In each dimension, $d$, 
 the mean of $B^{(d)}_k \phi$
 over sprinkling at density $\rho=l^{-d}$  takes the same form as for the original d'Alembertian but
with the discreteness scale $\rho$ replaced
 by $\rho_k = l_k^{-d}$. So, the mean, $\bar{B}^{(4)}_k\phi$,  of $B^{(4)}_k\phi$ is 
 given by (\ref{Bbar}) with $\rho$ replaced by $l_k^{-4}$.
Thus, results about $\bar{B}$ at finite $\rho$
can be applied to $\bar{B}_k$.
Simulations of $B_k^{(d)}$
 for a selection of test functions in 2, 3 and 4 dimensional flat space
 indicate that its fluctuations do die away as $l \rightarrow 0$ \cite{Sorkin:2007qi,
 Benincasa:2010ac,Dowker:2013vl}  but this remains to be more thoroughly tested.

In calculating the limit of the mean of the causal set d'Alembertian 
in curved space, we made the assumption that 
between $x$ and every point of 
$\partial J^-(x)$ there is a unique null geodesic.
This is a strong assumption and it is possible
that it can be weakened. The assumption is made so that $\partial J^-(x)$
can be treated as a null geodesic congruence, guaranteeing the 
existence of Null Gaussian Normal Coordinates in a neighbourhood of $\partial J^-(x)$.
When the assumption fails and there are caustics on $\partial J^-(x)$ it is nevertheless the case,
in a globally hyperbolic spacetime, that every point on $\partial J^-(x)$ lies on at least one
null geodesic from $x$. 
Moreover, the set of points on $\partial J^-(x)$ which are not connected to $x$ by a 
single null geodesic consists of those
points that lie on caustics and is a set of measure zero in $\partial J^-(x)$.
It might be possible to construct a proof in the general case by 
covering the region of integration down the light cone with appropriate finite 
collections of subregions in each of which Null Gaussian Normal Coordinates
can be constructed. If the integral can be performed in each subregion 
and shown to be equal to zero in the
limit, one might be able to argue that the whole integral also tends to zero.

The applicability of our result is limited by the fact that we have not
been able to estimate the finite $\rho$ corrections to the limit
 from the down-the-light-cone region in terms of physically interpretable 
 quantities. In order to 
estimate these corrections one needs an explicit asymptotic expansion
for the volume of causal intervals 
which hug the past light cone, the long skinny intervals.
These intervals have small volume but are not necessarily approximately flat. 
If, however,  
there is an approximately flat Riemann normal neighbourhood of the 
whole of $\partial J^-(x) \cap \textrm{Supp}(\phi)$ of thickness $a \gg l$, 
then $\bar{B}\phi(x)$ is a good
approximation to $\Box\phi(x)$ at finite density $\rho = l^{-4}$.
Moreover, if $\phi = 1$ and there is an approximately flat Riemann normal neighbourhood of the 
whole of $\partial J^-(x) \cap \textrm{Supp}(\phi)$ of thickness $a \gg l$, 
then $\bar{B}\phi(x)$ is a good
approximation to $-\frac{1}{2} R(x) $ at finite density $\rho = l^{-4}$. 
 
The value of the continuum limit of the mean of the causal set d'Alembertian
was used to propose a causal set action in $d=2,4$ \cite{Benincasa:2010ac} 
and in other dimensions \cite{Dowker:2013vl, Glaser:2013sf}.
Actions derived from $B_k$ have also been defined.
Our results suggest that the action in $d=4$ evaluated on a causal set that is a sprinkling 
of an approximately flat region of a four dimensional spacetime 
will be approximately equal to the Einstein-Hilbert (EH)
action of the spacetime -- if the fluctuations around the mean are 
small. We do not know what the value would be for a causal set that 
is a sprinkling into a spacetime which is large compared to the scale set by the 
curvature and it is possible it is \textit{not} close to the EH action. 
If this turns out to be the case, we would have a discrete action whose mean is close to the 
EH action for regions small compared to any curvature length scale, but starts to deviate
from it as the size of the region, $L$, approaches the curvature length scale.
If the proposed causal set action is relevant 
in a continuum regime of full quantum gravity, this might indicate that the 
continuum approximation to quantum gravity is General Relativity
for approximately flat regions of spacetime but deviates from it on scales 
large compared to the curvature scale.

\begin{acknowledgments}We thank Rafael Sorkin for helpful suggestions. 
FD  is supported by STFC grant ST/L00044X/1.
AB and DMTB acknowledge financial support from the John Templeton Foundation (JTF),  grant  No. 51876.
FD thanks the Aspen Center for Physics, Perimeter Institute, Waterloo, ON, Canada
and Institute for Quantum Computing, University of Waterloo, Waterloo, ON, Canada
for support. 
Research at Perimeter Institute is supported by the Government of Canada through Industry Canada and by the Province of Ontario through the Ministry of Economic Development and Innovation.\end{acknowledgments}

\bibliographystyle{apsrev}
\bibliography{refs}

\begin{thebibliography}{23}
\expandafter\ifx\csname natexlab\endcsname\relax\def\natexlab#1{#1}\fi
\expandafter\ifx\csname bibnamefont\endcsname\relax
  \def\bibnamefont#1{#1}\fi
\expandafter\ifx\csname bibfnamefont\endcsname\relax
  \def\bibfnamefont#1{#1}\fi
\expandafter\ifx\csname citenamefont\endcsname\relax
  \def\citenamefont#1{#1}\fi
\expandafter\ifx\csname url\endcsname\relax
  \def\url#1{\texttt{#1}}\fi
\expandafter\ifx\csname urlprefix\endcsname\relax\def\urlprefix{URL }\fi
\providecommand{\bibinfo}[2]{#2}
\providecommand{\eprint}[2][]{\url{#2}}

\bibitem[{\citenamefont{Sorkin}(1997)}]{Sorkin:1997gi}
\bibinfo{author}{\bibfnamefont{R.~D.} \bibnamefont{Sorkin}},
  \bibinfo{journal}{Int. J. Theor. Phys.} \textbf{\bibinfo{volume}{36}},
  \bibinfo{pages}{2759} (\bibinfo{year}{1997}), \eprint{gr-qc/9706002}.

\bibitem[{\citenamefont{Bombelli et~al.}(1987)\citenamefont{Bombelli, Lee,
  Meyer, and Sorkin}}]{Bombelli:1987aa}
\bibinfo{author}{\bibfnamefont{L.}~\bibnamefont{Bombelli}},
  \bibinfo{author}{\bibfnamefont{J.-H.} \bibnamefont{Lee}},
  \bibinfo{author}{\bibfnamefont{D.}~\bibnamefont{Meyer}}, \bibnamefont{and}
  \bibinfo{author}{\bibfnamefont{R.}~\bibnamefont{Sorkin}},
  \bibinfo{journal}{Phys. Rev. Lett} \textbf{\bibinfo{volume}{59}},
  \bibinfo{pages}{521} (\bibinfo{year}{1987}).

\bibitem[{\citenamefont{Moore}(1988)}]{Moore:1988zz}
\bibinfo{author}{\bibfnamefont{C.}~\bibnamefont{Moore}},
  \bibinfo{journal}{Phys. Rev. Lett.} \textbf{\bibinfo{volume}{60}},
  \bibinfo{pages}{655} (\bibinfo{year}{1988}).

\bibitem[{\citenamefont{Bombelli et~al.}(1988)\citenamefont{Bombelli, Lee,
  Meyer, and Sorkin}}]{Bombelli:1988qh}
\bibinfo{author}{\bibfnamefont{L.}~\bibnamefont{Bombelli}},
  \bibinfo{author}{\bibfnamefont{J.}~\bibnamefont{Lee}},
  \bibinfo{author}{\bibfnamefont{D.}~\bibnamefont{Meyer}}, \bibnamefont{and}
  \bibinfo{author}{\bibfnamefont{R.~D.} \bibnamefont{Sorkin}},
  \bibinfo{journal}{Phys. Rev. Lett.} \textbf{\bibinfo{volume}{60}},
  \bibinfo{pages}{656} (\bibinfo{year}{1988}).

\bibitem[{\citenamefont{Daughton}(1993)}]{Daughton:1993}
\bibinfo{author}{\bibfnamefont{A.}~\bibnamefont{Daughton}}, Ph.D. thesis,
  \bibinfo{school}{Syracuse University} (\bibinfo{year}{1993}).

\bibitem[{\citenamefont{Salgado}(2008)}]{Salgado:2008}
\bibinfo{author}{\bibfnamefont{R.~B.} \bibnamefont{Salgado}}, Ph.D. thesis,
  \bibinfo{school}{Syracuse University} (\bibinfo{year}{2008}).

\bibitem[{\citenamefont{Sorkin}(2006)}]{Sorkin:2007qi}
\bibinfo{author}{\bibfnamefont{R.~D.} \bibnamefont{Sorkin}}, in
  \emph{\bibinfo{booktitle}{{Approaches to Quantum Gravity: Towards a New
  Understanding of Space and Time}}}, edited by
  \bibinfo{editor}{\bibfnamefont{D.}~\bibnamefont{Oriti}}
  (\bibinfo{publisher}{Cambridge University Press}, \bibinfo{year}{2006}),
  \eprint{gr-qc/0703099}.

\bibitem[{\citenamefont{Henson}(2006)}]{Henson:2006kf}
\bibinfo{author}{\bibfnamefont{J.}~\bibnamefont{Henson}}, in
  \emph{\bibinfo{booktitle}{{Approaches to Quantum Gravity: Towards a New
  Understanding of Space and Time}}}, edited by
  \bibinfo{editor}{\bibfnamefont{D.}~\bibnamefont{Oriti}}
  (\bibinfo{publisher}{Cambridge University Press}, \bibinfo{year}{2006}),
  \eprint{gr-qc/0601121}.

\bibitem[{\citenamefont{Sorkin}()}]{Sorkinnotes}
\bibinfo{author}{\bibfnamefont{R.}~\bibnamefont{Sorkin}},
  \bibinfo{howpublished}{unpublished notes, private communication}.

\bibitem[{\citenamefont{Benincasa and Dowker}(2010)}]{Benincasa:2010ac}
\bibinfo{author}{\bibfnamefont{D.~M.~T.} \bibnamefont{Benincasa}}
  \bibnamefont{and} \bibinfo{author}{\bibfnamefont{F.}~\bibnamefont{Dowker}},
  \bibinfo{journal}{Phys. Rev. Lett.} \textbf{\bibinfo{volume}{104}},
  \bibinfo{pages}{181301} (\bibinfo{year}{2010}),
  \eprint{http://arxiv.org/abs/1001.2725}.

\bibitem[{\citenamefont{Dowker and Glaser}(2013)}]{Dowker:2013vl}
\bibinfo{author}{\bibfnamefont{F.}~\bibnamefont{Dowker}} \bibnamefont{and}
  \bibinfo{author}{\bibfnamefont{L.}~\bibnamefont{Glaser}},
  \bibinfo{journal}{2013,Class. Quantum Grav. 30 195016}
  (\bibinfo{year}{2013}), \eprint{1305.2588},
  \urlprefix\url{http://arxiv.org/abs/1305.2588}.

\bibitem[{\citenamefont{Glaser}(2014)}]{Glaser:2013sf}
\bibinfo{author}{\bibfnamefont{L.}~\bibnamefont{Glaser}},
  \bibinfo{journal}{Classical and Quantum Gravity}
  \textbf{\bibinfo{volume}{31}}, \bibinfo{pages}{095007}
  (\bibinfo{year}{2014}),
  \urlprefix\url{http://stacks.iop.org/0264-9381/31/i=9/a=095007}.

\bibitem[{\citenamefont{Belenchia}(2016)}]{0264-9381-33-13-135011}
\bibinfo{author}{\bibfnamefont{A.}~\bibnamefont{Belenchia}},
  \bibinfo{journal}{Classical and Quantum Gravity}
  \textbf{\bibinfo{volume}{33}}, \bibinfo{pages}{135011}
  (\bibinfo{year}{2016}),
  \urlprefix\url{http://stacks.iop.org/0264-9381/33/i=13/a=135011}.

\bibitem[{\citenamefont{Aslanbeigi et~al.}(2014)\citenamefont{Aslanbeigi,
  Saravani, and Sorkin}}]{Aslanbeigi:2014tg}
\bibinfo{author}{\bibfnamefont{S.}~\bibnamefont{Aslanbeigi}},
  \bibinfo{author}{\bibfnamefont{M.}~\bibnamefont{Saravani}}, \bibnamefont{and}
  \bibinfo{author}{\bibfnamefont{R.~D.} \bibnamefont{Sorkin}},
  \bibinfo{journal}{JHEP} \textbf{\bibinfo{volume}{1406}}, \bibinfo{pages}{024}
  (\bibinfo{year}{2014}), \eprint{1403.1622},
  \urlprefix\url{http://arxiv.org/abs/1403.1622}.

\bibitem[{\citenamefont{Belenchia et~al.}(2015)\citenamefont{Belenchia,
  Benincasa, and Liberati}}]{Belenchia:2014fda}
\bibinfo{author}{\bibfnamefont{A.}~\bibnamefont{Belenchia}},
  \bibinfo{author}{\bibfnamefont{D.~M.~T.} \bibnamefont{Benincasa}},
  \bibnamefont{and} \bibinfo{author}{\bibfnamefont{S.}~\bibnamefont{Liberati}},
  \bibinfo{journal}{JHEP} \textbf{\bibinfo{volume}{1503}}, \bibinfo{pages}{036}
  (\bibinfo{year}{2015}), \eprint{1411.6513}.

\bibitem[{\citenamefont{Saravani and Aslanbeigi}(2015)}]{Saravani:2015aa}
\bibinfo{author}{\bibfnamefont{M.}~\bibnamefont{Saravani}} \bibnamefont{and}
  \bibinfo{author}{\bibfnamefont{S.}~\bibnamefont{Aslanbeigi}}
  (\bibinfo{year}{2015}), \eprint{1502.01655},
  \urlprefix\url{http://arxiv.org/abs/1502.01655}.

\bibitem[{\citenamefont{Saravani and Afshordi}(2016)}]{Saravani:2016aa}
\bibinfo{author}{\bibfnamefont{M.}~\bibnamefont{Saravani}} \bibnamefont{and}
  \bibinfo{author}{\bibfnamefont{N.}~\bibnamefont{Afshordi}}
  (\bibinfo{year}{2016}), \eprint{1604.02448},
  \urlprefix\url{http://arxiv.org/abs/1604.02448}.

\bibitem[{\citenamefont{Belenchia et~al.}(2016)\citenamefont{Belenchia,
  Benincasa, Martin-Martinez, and Saravani}}]{Belenchia:2016aa}
\bibinfo{author}{\bibfnamefont{A.}~\bibnamefont{Belenchia}},
  \bibinfo{author}{\bibfnamefont{D.~M.~T.} \bibnamefont{Benincasa}},
  \bibinfo{author}{\bibfnamefont{E.}~\bibnamefont{Martin-Martinez}},
  \bibnamefont{and} \bibinfo{author}{\bibfnamefont{M.}~\bibnamefont{Saravani}}
  (\bibinfo{year}{2016}), \eprint{1605.03973},
  \urlprefix\url{http://arxiv.org/abs/1605.03973}.

\bibitem[{\citenamefont{Hawking and Ellis}(1973)}]{hawking1973large}
\bibinfo{author}{\bibfnamefont{S.~W.} \bibnamefont{Hawking}} \bibnamefont{and}
  \bibinfo{author}{\bibfnamefont{G.~F.~R.} \bibnamefont{Ellis}},
  \emph{\bibinfo{title}{The large scale structure of space-time}},
  vol.~\bibinfo{volume}{1} (\bibinfo{publisher}{Cambridge university press},
  \bibinfo{year}{1973}).

\bibitem[{\citenamefont{Friedrich et~al.}(1999)\citenamefont{Friedrich, Racz,
  and Wald}}]{Friedrich:1999aa}
\bibinfo{author}{\bibfnamefont{H.}~\bibnamefont{Friedrich}},
  \bibinfo{author}{\bibfnamefont{I.}~\bibnamefont{Racz}}, \bibnamefont{and}
  \bibinfo{author}{\bibfnamefont{R.~M.} \bibnamefont{Wald}},
  \bibinfo{journal}{Commun.Math.Phys.} \textbf{\bibinfo{volume}{204}},
  \bibinfo{pages}{691} (\bibinfo{year}{1999}), \eprint{gr-qc/9811021},
  \urlprefix\url{http://arxiv.org/abs/gr-qc/9811021}.

\bibitem[{\citenamefont{Myrheim}(1978)}]{Myrheim:1978}
\bibinfo{author}{\bibfnamefont{J.}~\bibnamefont{Myrheim}},
  \emph{\bibinfo{title}{Statistical geometry}}, \bibinfo{howpublished}{CERN
  preprint TH-2538} (\bibinfo{year}{1978}).

\bibitem[{\citenamefont{Gibbons and Solodukhin}(2007)}]{Gibbons:2007aa}
\bibinfo{author}{\bibfnamefont{G.~W.} \bibnamefont{Gibbons}} \bibnamefont{and}
  \bibinfo{author}{\bibfnamefont{S.~N.} \bibnamefont{Solodukhin}},
  \bibinfo{journal}{Phys.Lett.B} \textbf{\bibinfo{volume}{649}},
  \bibinfo{pages}{317} (\bibinfo{year}{2007}), \eprint{hep-th/0703098},
  \urlprefix\url{http://arxiv.org/abs/hep-th/0703098}.

\bibitem[{\citenamefont{Blau et~al.}(2006)\citenamefont{Blau, Frank, and
  Weiss}}]{Blau:2006aa}
\bibinfo{author}{\bibfnamefont{M.}~\bibnamefont{Blau}},
  \bibinfo{author}{\bibfnamefont{D.}~\bibnamefont{Frank}}, \bibnamefont{and}
  \bibinfo{author}{\bibfnamefont{S.}~\bibnamefont{Weiss}},
  \bibinfo{journal}{Class.Quant.Grav.} \textbf{\bibinfo{volume}{23}},
  \bibinfo{pages}{3993} (\bibinfo{year}{2006}), \eprint{hep-th/0603109},
  \urlprefix\url{http://arxiv.org/abs/hep-th/0603109}.

\end{thebibliography}

\appendix
\section{}
\label{dlcvolume}

Consider a point $y$ in $N_{LC}$ with Null Gaussian Normal Coordinates $(U, V, \theta, \phi)$ and 
the volume, $V(y)$ of the causal interval, Int$(x,y)$, between $x$ and $y$.

We use the null geodesic $\gamma(\theta,\varphi)$ defined in the text to 
define  Null Fermi Normal Coordinates  (NFNC) $(x^+, x^-, x^1, x^2)$ associated to $\gamma(\theta, \varphi)$
\cite{Blau:2006aa}. They are defined using a pseudo-orthonormal tetrad at $x$,
$\{E_+ , E_- , E_1, E_2\}$ where $E_+ = T(\theta, \varphi)$, $E_- = T(\pi -\theta, \varphi+\pi)$ and
$E_1$ and $E_2$ are spacelike unit vectors, orthogonal to each other and to 
both $T(\theta, \varphi)$ and $T(\pi -\theta,
\varphi+ \pi)$. The 
affine parameter along $\gamma(\theta,\varphi)$ is $x^+$.  
This tetrad is parallel transported along $\gamma(\theta,\varphi)$: 
$\{E_+(x^+) , E_-(x^+) , E_1(x^+), E_2(x^+)\}$.
The NFNC are normal coordinates defined by the spray of geodesics 
emanating from each point along $\gamma(\theta,\varphi)$ with 
tangent vectors lying in the subspace of the tangent space
spanned by $\{E_- (x^+), E_1(x^+), E_2(x^+)\}$.

In these NFNC the point $y$ has coordinates $(x^+ = V, x^- = U, x^1= 0, x^2 = 0)$. 
Indeed,  the 2-dimensional surface defined in NGNC by $\theta$ constant and 
$\phi$ constant is the same surface as that defined in the NFNC (associated with
$\gamma(\theta, \varphi)$)  by $x^1= x^2 =0$, where the two coordinate systems 
overlap. 

In NFNC the metric to quadratic order is \cite{Blau:2006aa}
\begin{align}
ds^2&= -2dx^+dx^- +\delta_{ab} dx^a dx^b \nonumber\\
&-\left[R_{+\bar{a}+{\bar{b}}}(x^+) \ x^{\bar{a}} x^{\bar{b}} (dx^+)^2 
         +\frac{4}{3} R_{+{\bar{b}}\bar{a}{\bar{c}}} (x^+)
          x^{\bar{b}} x^{\bar{c}} (dx^+ dx^{\bar{a}}) +\frac{1}{3}
          R_{\bar{a}{\bar{c}}{\bar{b}}{\bar{d}}} (x^+) x^{\bar{c}} x^{\bar{d}}
          (dx^{\bar{a}} dx^{\bar{b}})\right]\nonumber\\
   &+ \mathcal{O}(x^{\bar{a}}x^{\bar{b}}x^{\bar{c}})
\label{metex}
\end{align}
where all the curvature components are evaluated on the null geodesic, 
the barred indices, $\bar{a}, \bar{b}$ {\textit{etc.}}, run over the three transverse 
directions $-, 1,2$ and unbarred, $a,b$ {\textit{etc.}},
 over the spatial transverse directions  $1$ and $2$ 
only. Note: there is a sign difference between the $x^-$ coordinate of Blau et al and 
our $x^-$ coordinate here.

Fixing $y$ with its NGNC $\{U, V, \theta, \phi\}$, we assume that the causal interval 
Int$(x,y)$ between $x$ and $y$ lies in the tubular neighbourhood of $\gamma(\theta,\varphi)$ 
on which the NFNC are defined. We want to evaluate 
\begin{equation}
V(y) = \int_{\textrm{Int}(x,y)} dx^+ dx^- dx^1 dx^2 \sqrt{-g( x^+, x^-, x^1, x^2)}\,,\label{volskinny}
\end{equation}
the volume of Int$(x,y)$  as an expansion 
in $U$ as $U\rightarrow 0$. In other words we want to consider the limit as the interval
tends to the segment of the null geodesic $\gamma(\theta, \varphi)$ between 
$x$ and the point with NGNC $\{0, V, \theta, \phi\}$. This is related to the Penrose limit. 

Rescaling the coordinates 
$z^-=x^-/U$, $z^+ = x^+$  and $ z^a=x^a/\sqrt{U}$, in the limit $U\rightarrow0$ the metric becomes
\begin{align}
ds^2 = U\left(-2dz^+dz^- +{\delta_{ab}}dz^a dz^b - R_{+a+b}(z^+)
z^a z^b(dz^+)^2  \right) + O(U^{\frac{3}{2}})
\end{align}
and the next terms proportional to $U^{\frac{3}{2}}$ and $U^{2}$ in the expansion can be found in 
Appendix A of reference \cite{Blau:2006aa}.
Using this one can show that $\sqrt{-g( z^+, z^-, z^1, z^2)} = U^2[1 + U f(U, z^+, z^-, z^1, z^2)]  $, 
where, we assume, $f$ is a differentiable function of $U$. 
\begin{align}
V(y) = &\,\int_{\textrm{Int}(x,y)} dz^+ dz^- dz^1 dz^2 \sqrt{-g( z^+, z^-, z^1, z^2)}\\
= &\, U^2 \int_{\textrm{Int}(x,y)} dz^+ dz^- dz^1 dz^2 [1 + U f(U, z^+, z^-, z^1, z^2)]\,.
\label{volskinny}
\end{align}

The interval $\textrm{Int}(x,y)$ is defined by the causal structure of the metric and 
is the same for $ds^2$ as for the conformally rescaled metric
\begin{align}
{\widetilde{ds}}^2 = U^{-1}ds^2 = -2dz^+dz^- +{\delta_{ab}}dz^a dz^b - R_{+a+b}(z^+)
z^az^b(dz^+)^2   + O(U^{\frac{1}{2}})\,.
\end{align}
In the $U\rightarrow 0$ limit, therefore, the integral 
\begin{equation}
\int_{\textrm{Int}(x,y)} dz^+ dz^- dz^1 dz^2
\label{volpenrose}
\end{equation}
tends to the volume of the causal interval between the origin and 
the point with coordinates $(z^+ = V, z^- = 1, z^1 = 0, z^2 = 0)$ in the 
Penrose limit metric
\begin{align}\label{penroselim}
ds^2 = -2dz^+dz^- +{\delta_{ab}}dz^a dz^b - R_{+a+b}(z^+)
z^az^b(dz^+)^2 \,.
\end{align}
Note that $R_{+a+b}(z^+)$ are the curvature components of the original metric
along $\gamma(\theta, \phi)$ in the original, unscaled NFNC. 
We denote this limit volume by $f_0(V, \theta, \phi)$. 
It is an open question whether $f_0$ can be expressed more concretely
 in terms of (integrals of?) the curvature components along $\gamma(\theta, \phi)$.

We conclude that $U^{-2} V(y) \rightarrow f_0(V, \theta, \phi)$ as $U\rightarrow 0$ and 
so $V(y) = U^2 f_0(V, \theta, \phi) + U^3 G(V, U, \theta, \varphi)$ 
with $G$ a continuous function of $U$. In fact, we need to assume more differentiability 
than this for $V(y)$ for the proof but the crucial fact established here is the 
$U^2$ behaviour of the leading term in the expansion. 

We can also show that $f_0$ is a monotonic increasing function of $V$. 
Consider two points $p$ and $p'$ in the Penrose limiting geometry (\ref{penroselim})
with coordinates $(z^+ = V, z^- = 1, z^1 = 0, z^2 = 0)$ and 
$(z^+ = V', z^- = 1, z^1 = 0, z^2 = 0)$ 
respectively,  where $V<V'$. There is a future pointing null geodesic, along 
which $z^- = 1, z^1 = 0, z^2 = 0$, from $p'$ to $p$ and so the causal 
interval from $p'$ to $x$ contains the causal interval from $p$ to $x$.
Then, if $U$ is small enough,  it follows
that $V(y)$ is monotonic increasing in $V$.

\end{document}